\documentclass[journal]{IEEEtran}
\usepackage{graphicx}
\usepackage{booktabs}
\usepackage{subfigure}
\usepackage{overpic}
\usepackage{amsmath}
\usepackage{flushend}
\usepackage{amssymb}
\usepackage{multirow}
\usepackage{makecell}
\usepackage{listings}
\usepackage{algorithm}
\usepackage{fontawesome}
\usepackage{color, soul}
\usepackage[absolute,overlay]{textpos}                    

\usepackage{hyperref}
\hypersetup{
	colorlinks=true,
	linkcolor=cyan,
	urlcolor=magenta,
	citecolor=cyan,
}
\usepackage{colortbl}
\usepackage[dvipsnames]{xcolor}
\definecolor{bg}{HTML}{e0f1ff}

\begin{document}

\title{Nearest Neighbor-Based Contrastive Learning for Hyperspectral and LiDAR Data Classification}
\author{Meng Wang, Feng Gao, Junyu Dong, Heng-Chao Li, Qian Du 
\thanks{This work was supported in part by the National Key Research and Development Program of China under Grant 2018AAA0100602, and in part by the National Natural Science Foundation of China under Grant 42106191.

Meng Wang, Feng Gao, and Junyu Dong are with the School of Information Science and Engineering, Ocean University of China, Qingdao 266100, China. \emph{(Corresponding author: Feng Gao.)}

H. -C. Li is with the Sichuan Provincial Key Laboratory of Information Coding and Transmission, Southwest Jiaotong University, Chengdu 610031, China.

Qian Du is with the Department of Electrical and Computer Engineering, Mississippi State University, Starkville, MS 39762 USA.}}

\markboth{IEEE TRANSACTIONS ON GEOSCIENCE AND REMOTE SENSING}%
{Shell}

\maketitle

\begin{abstract}
The joint hyperspectral image (HSI) and LiDAR data classification aims to interpret ground objects at more detailed and precise level. Although deep learning methods have shown remarkable success in the multisource data classification task, self-supervised learning has rarely been explored. It is commonly nontrivial to build a robust self-supervised learning model for multisource data classification, due to the fact that the semantic similarities of neighborhood regions are not exploited in existing contrastive learning framework. Furthermore, the heterogeneous gap induced by the inconsistent distribution of multisource data impedes the classification performance. To overcome these disadvantages, we propose a \underline{N}earest \underline{N}eighbor-based \underline{C}ontrastive Learning \underline{Net}work (NNCNet), which takes full advantage of large amounts of unlabeled data to learn discriminative feature representations. Specifically, we propose a nearest neighbor-based data augmentation scheme to use enhanced semantic relationships among nearby regions. The intermodal semantic alignments can be captured more accurately. In addition, we design a bilinear attention module to exploit the second-order and even high-order feature interactions between the HSI and LiDAR data. Extensive experiments on four public datasets demonstrate the superiority of our NNCNet over state-of-the-art methods. The source codes are available at \url{https://github.com/summitgao/NNCNet}.
\end{abstract}


\IEEEpeerreviewmaketitle

\section{Introduction}

\IEEEPARstart{R}{ecently}, with the rapid development of satellite sensors, an ever increasing number of multimodal images (optical, SAR, hyperspectral and LiDAR) are obtained everyday \cite{ma21tgrs}. Among these multimodal data, hyperspectral images (HSIs) provide detailed spectral information for the identification of specified objects on the ground, while LiDAR data provide elevation information of the area \cite{khod15jstars} \cite{rasti17tgrs} \cite{zheng22tip}. These HSI and LiDAR sensors are different in imaging mechanism, spatial resolution, and even coverage. Therefore, both sensors capture different properties of the earth, such as spectral radiance and height information. For example, there are no significant differences in the spectral domain between the ``trees" on the ground and the ``trees" on the hill, but they can be distinguished from the LiDAR data \cite{hong21tgrs}. Therefore, the joint exploitation of HSI and LiDAR data enables us to interpret ground objects at a more detailed and precise level, which can hardly be achieved by using single-mode data \cite{chova15}. Thus, the classification of cross-modal data has attracted considerable attention and has been widely applied in multisource image interpretations \cite{ge19jstars} \cite{graph22tgrs}.

\begin{figure}
    \centering
    \includegraphics[width=3.4in]{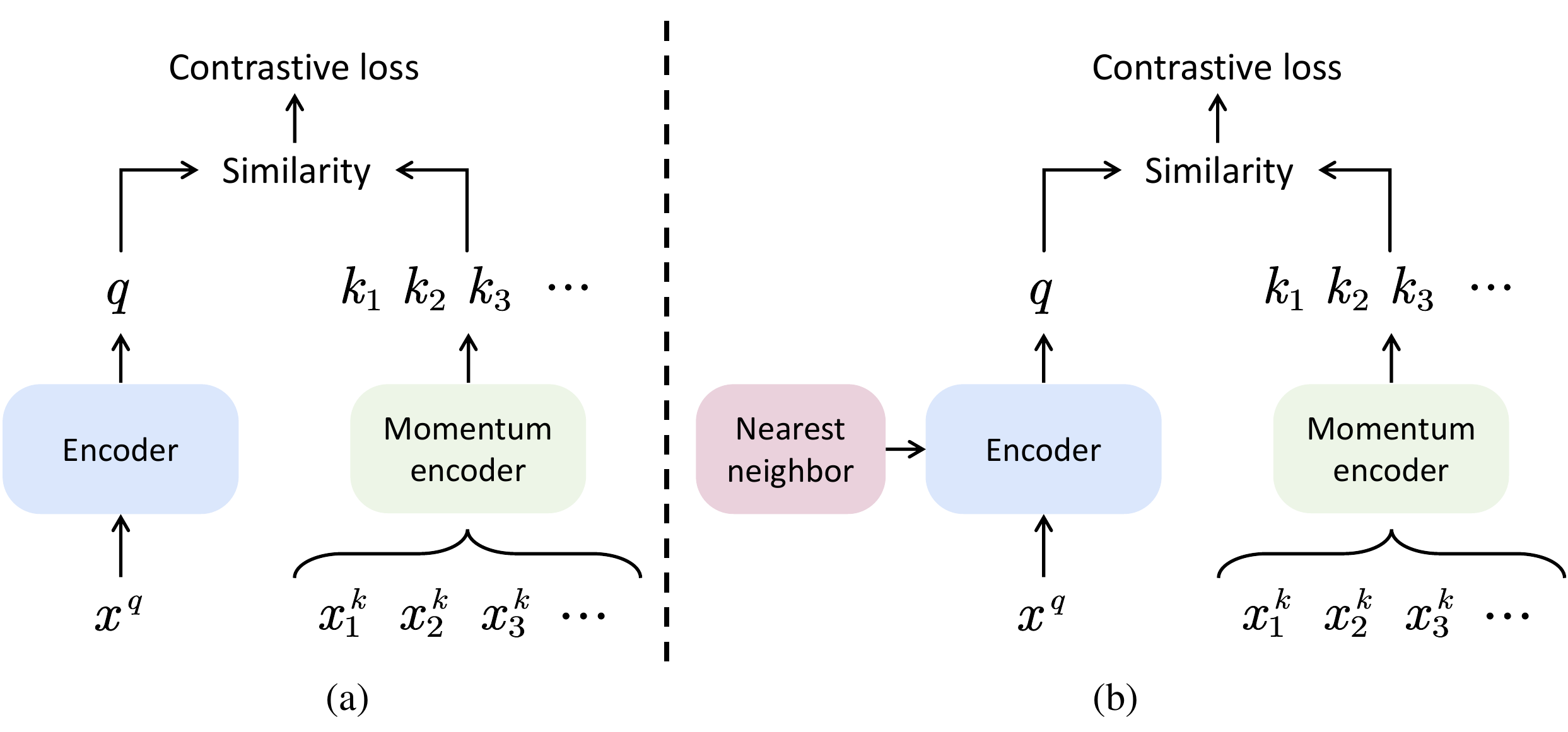}
    \caption{Conceptual comparison of MoCo and the proposed nearest neighbor-based contrastive learning framework. In the proposed framework, the nearest neighbors are considered as positive samples. The semantic similarities among neighborhood regions are exploited.}
    \label{fig_comp}
\end{figure}

A great deal of effort has been put into solving the problem of HSI and LiDAR joint classification. Traditionally, feature-level fusion models have been proposed, and these models commonly concatenate the HSI and LiDAR features for classification \cite{pedergnana12jstars} \cite{huang13igarss} \cite{demirkesen14igarss}. Besides feature-level fusion, decision-level fusion is another popular solution for HSI and LiDAR classification. Several classifiers are designed for HSI and LiDAR data, respectively. The voting strategy is commonly used to obtain the final classification map \cite{xia17icassp}. Subsequently, to further exploit high-level semantic features, convolutional neural networks (CNNs) are employed for multisource data classification \cite{zhang21tgrs}. Encoder-decoder network \cite{endnet}, coupled CNNs \cite{hang20tgrs}, Gabor CNN \cite{zhao21tgrs}, cross attention \cite{gao21tgrs}, and Transformer \cite{tranformer22tip} are used to extract representative multisource features, and these methods have achieved promising performance.

In practice, deep learning models have demonstrated remarkable success in various multisource data joint classification. However, it is non-trivial to build an effective HSI and LiDAR classification model. One of the critical reasons is that the deep learning-based model commonly requires a great number of labeled samples to achieve satisfactory accuracy, which is expensive and limited in ground object modeling. Recent research in self-supervised learning encourages the deep network to learn more representative and interpretable features in natural language processing \cite{sarkar19taslp} \cite{liu21taslp} and computer vision tasks \cite{xu21tpami} \cite{he20cvpr}. Self-supervised learning mines the inherent attributes and semantics of large-scale unlabeled data to obtain beneficial semantic representations, and it does not require manually annotated data \cite{jing20tpami}. After the self-supervised training finished, the learned  features can be transferred to classification tasks (especially when only small training data is available) as pretrained models to boost the classification performance and alleviate overfitting \cite{ren21tgrs} \cite{jung21grsl} \cite{yue21tgrs} \cite{zheng22smalldata}. In HSI and LiDAR joint classification, self-supervised learning has rarely been explored, and in this paper, we aim to build an effective self-supervised model to solve the problem.

It is commonly non-trivial to build a robust self-supervised learning model for the HSI and LiDAR joint classification task, due to the following reasons: \textbf{1) Data augmentation scheme.} In Momentum Contrast (MoCo) for self-supervised learning \cite{he20cvpr}, the random color jittering, random horizontal flip, and random grayscale conversion are used for data augmentation. However, such data augmentation scheme does not take the spatial distances between the positive and negative samples into account, and the semantic similarities of neighborhood regions are not exploited. Consequently, how to properly utilize the semantic similarities among nearby regions is a major challenge. \textbf{2) The heterogeneous gap.} HSI and LiDAR joint classification requires a comprehensive understanding of complex heterogeneous data simultaneously. However, the heterogeneous gap induced by the inconsistent distributions of multisource data would greatly impedes its implementation. Therefore, it is vital to bridge this gap for more robust multisource data classification. 

To address the aforementioned challenges, we propose a \textbf{N}earest \textbf{N}eighbor-based \textbf{C}ontrast learning \textbf{Net}work, \textbf{NNCNet} for short, which aims to learn an encoder that encodes similar data of the same kind and makes the encoding results of different classes of data as different as possible. To be more specific, we propose a nearest neighbor-based framework to use the enhanced semantic relationships among nearby regions. As illustrated in Fig. \ref{fig_comp}, nearest neighbors of positive samples are fed into the encoder for contrastive learning. The feature representations are learned by encouraging the proximity of between different views of the same sample and its nearest neighbors in the spatial domain. Therefore, the contrastive learning framework is encouraged to generalize to new feature embeddings that may not be covered by the data augmentation. In addition, we design a bilinear attention fusion module to exploit second-order and even higher-order feature interactions between the HSI and LiDAR data, and the information flow can be controlled more flexibly. 

The contributions of this work are as follows:

\begin{itemize}

\item We propose a self-supervision contrastive learning approach NNCNet, which integrates a nearest neighbor-based data augmentation scheme. The scheme can exploit the semantic similarities among neighborhood regions, and hence capture inter-modal semantic alignments more accurately. To our best knowledge, we are the first to apply self-supervised contrastive learning to HSI and LiDAR joint classification, which has both great theoretical and practical significance.

\item We propose a bilinear attention fusion module that aims to enhance the contextual representation of HSI and LiDAR data. The module captures second-order feature interactions between multisource data.

\item We have conducted extensive experiments on four benchmark datasets to validate the effectiveness of our NNCNet. Additionally, we have released our codes and parameters to benefit other researchers.

\end{itemize}

\section{Related Work}

\subsection{Morphological Filter-Based Methods for HSI and LiDAR Classification}

The joint use of HSI and LiDAR has already been investigated for a variety of applications, such as illumination calibration \cite{brell17tgrs}, forest area analysis \cite{dalponte08tgrs}, bushfire monitoring \cite{koetz08fem}, and urban sprawl modeling \cite{heiden12lup}. Great efforts have been devoted to exploiting the complementary information between multisource data, especially for morphological filter-based methods. Morphological filters are intensively used to attenuate the redundant spatial details and preserve the geometric structures. Pedergnana et al. \cite{pedergnana12jstars} used morphological extended attribute profiles to HSI and LiDAR data for classification. Features extracted from HSI and LiDAR data are stacked for classification.  Liao et al. \cite{liao15grsl} computed morphological attribute profiles from HSI and LiDAR data, and these attribute profiles are fused using a generalized graph-based method.  Khodadadzadeh et al. \cite{khodadadzadeh15jstars} pointed out that simple stacking of morphological attribute profiles from multisource data may contain redundant features. To solve this issue, they proposed a multiple feature learning approach based on the multinomial logistic regression classifier, which can adaptively exploit the spatially and spectrally derived features. Later, attribute profiles are considered to be complex and time-consuming in threshold initialization, and extinction profiles \cite{ghamisi16tgrs} are proposed to solve the problem. Ghamisi et al. \cite{ghamisi17jstars} presented a classification framework based on extinction profiles and deep learning.

\subsection{CNN-Based Methods for HSI and LiDAR Classification}

Recently, deep CNNs have attracted extensive research attention in the remote sensing data fusion community, and many CNN-based models have been proposed for multisource data classification. Xu et al. \cite{xu18tgrs} proposed a two-branch CNN model, which consists of a 2-D convolutional network and a 1-D convolutional network. Zhang et al. \cite{zhang20tcyb} presented a patch-to-patch CNN for the joint feature extraction of HSI and LiDAR data. Chen et al. \cite{chen17grsl} proposed a CNN and DNN hybrid model for multisource feature extraction. CNNs are used to extract informative features from multisource data, and a DNN is utilized to fuse these heterogeneous features for robust classification.  Li et al. \cite{li20tnnls} proposed a dual-channel spatial, spectral and multiscale attention CNN for multisource data classification. Hang et al. \cite{hang20tgrs} used coupled CNNs for multisource data classification. The coupled layers reduce the number of parameters and guide both networks learning from each other. Zhao et al. \cite{zhao21tgrs} proposed a fractional Gabor CNN, and focused on efficient feature fusion. Fractional Gabor convolutional kernels are used for multiscale and multidirectional feature extraction and yield robust feature representations against semantic changes. In \cite{du21tgrs}, a multisource graph fusion network is presented to integrate feature extraction and fusion into a single network. A multimodal graph is constructed to guide the multimodal image feature extraction. Gao et al. \cite{gao21tgrs} proposed a deep-wise feature interaction network for multisource remote sensing image classification. Consistency loss, discrimination loss, and classification loss are designed for parameter optimization.

\begin{figure*}[htb]
    \centering
    \includegraphics[width=.9\textwidth]{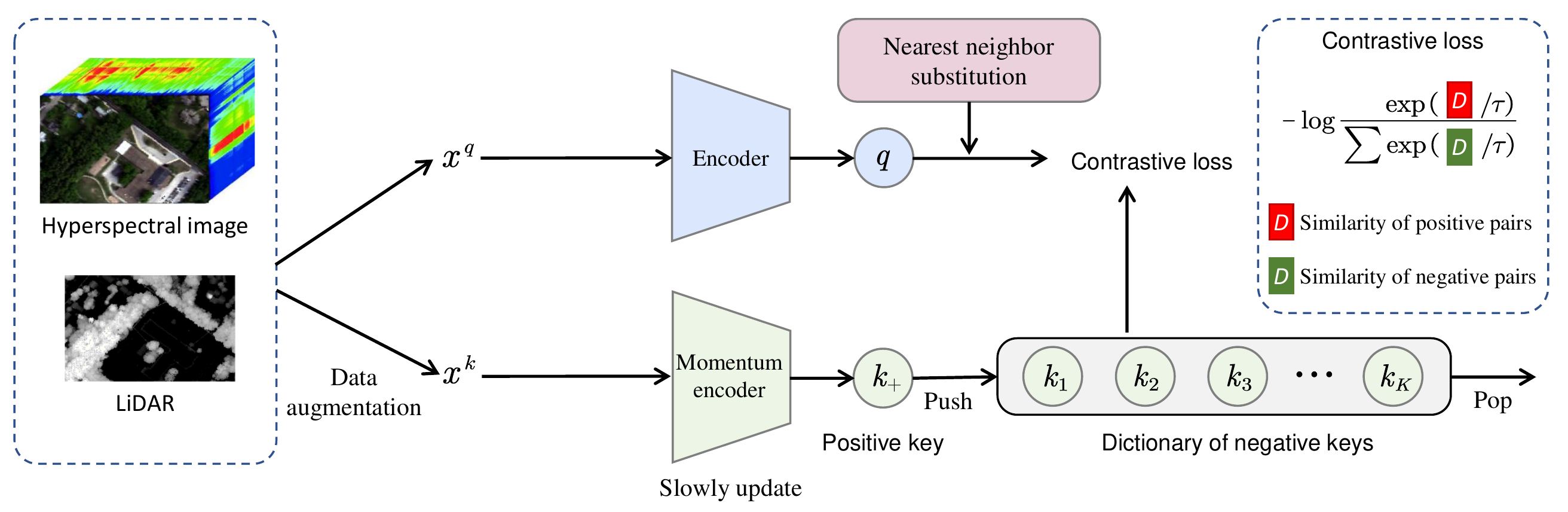}
    \caption{Schematic illustration of the nearest-neighbor based contrastive learning. It consists of three components: 1) Nearest neighbor-based data augmentation. The input samples are handled by random data augmentation to generate query and key samples. In a mini-batch, half positive key samples are substituted by its nearest neighbors to form new positive key samples. These nearest neighbors act small semantic perturbations. 2) Bilinear attention-based feature encoder. The query and key samples are fed into the encoder for feature extraction. A bilinear attention fusion module is employed to capture the second-order feature interactions between multisource data. 3) Contrastive loss computation. Positive and negative keys are stored in a dynamic dictionary, and contrastive loss is computed to assign high scores for positive keys and low scores for negative keys.}
    \label{fig_frame}   
\end{figure*} 

\begin{figure}
    \centering
    \includegraphics[width=3in]{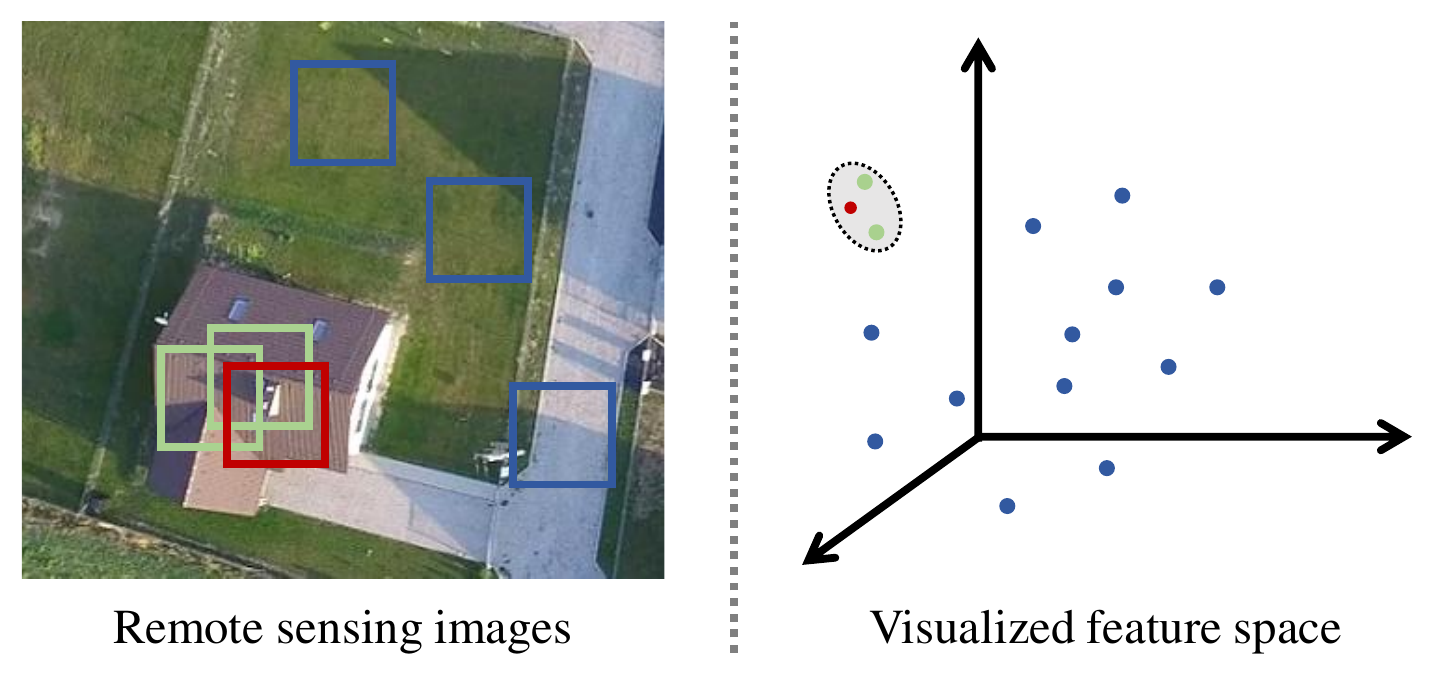}
    \caption{Typical regions in remote sensing images and the corresponding visualized features. The region within the red box is rooftop, and the green boxes are its nearest neighbors. Blue boxes denote regions far from the the red box. In the visualized feature space, it can be observed that features of the nearest neighbors are close. Therefore, the contextual information is critical in contrastive learning for remote sensing image classification.}
    \label{fig_nnvis}
\end{figure}

Although CNNs have been successfully applied to HSI and LiDAR joint classification, its performance remains unsatisfactory for practical applications. An important factor may be the lack of sufficient annotated data. In practical applications, remote sensing data annotation is costly, making it difficult to obtain robust deep learning models. To solve this problem, we aim to build a simple yet effective self-supervised method for multisource data joint classification. It extracts the inherent attributes and semantics from unlabeled large-scale data to capture beneficial feature representations. In addition, a nearest-neighbor-based data augmentation scheme is used to exploit the semantic relationships among nearby regions.

\section{Methodology}

As shown in Fig. \ref{fig_frame}, the proposed NNCNet consists of three parts: nearest neighbor-based data augmentation, bilinear attention-based encoder, and contrastive loss computation. Considering that the proposed NNCNet is based on a self-supervised contrastive learning framework, we first introduce the nearest neighbor-based contrastive learning framework and then successively elaborate the bilinear attention-based encoder.

\subsection{Nearest Neighbor-based Momentum Contrast Learning Framework}

Considering that unlabeled data have no supervised information, we aim to extract the supervised information from large-scale unsupervised HSI and LiDAR data. Our goal is to train an encoder that keeps the different transformations from the same sample as close as possible and the different samples as far away as possible in the feature space. To solve the problem, He et al. \cite{he20cvpr} proposed Momentum Contrast (MoCo) for self-supervised learning. To be specific, a minibatch of samples is selected from the data. Each sample is handled by random data augmentation (Gaussian blur, flip, or spectral distortions) to generate a query sample and a key sample. The query and key samples are encoded separately to embedding $q$ and $k$. The cosine similarity between $q$ and $k$ is computed for representation learning. The embedding from the same image is defined as the positive key, and embedding from different image is defined as the negative key. In MoCo, a dynamic dictionary is built with a queue and a dynamic encoder. 

For remote sensing data classification, we argue that random augmentations can hardly provide positive pairs for the same object representation. For the sake of covering more variance in a given class, we propose nearest neighbor-based contrastive learning framework. In remote sensing images, the neighbor labels of one specified position tend to be the same. As illustrated in Fig. \ref{fig_nnvis}, the region within the red box is rooftop, and the green boxes are its nearest neighbors. Blue boxes denote regions far from the red box. From the visualized feature space, it can be observed that the features of the nearest neighbors are close. In this paper, we use a nearest neighbor-based contrastive learning framework in which the semantic similarities among neighborhood regions are exploited. Therefore, the inter-modal semantic alignments are reinforced.

\begin{algorithm}[t]
\caption{Pseudocode of the Nearest Neighobr-Based Contrastive Learning in PyTorch Style.}
\label{alg_code}
\definecolor{col}{rgb}{0.3,0.75,0.3}
\lstset{
  backgroundcolor=\color{white},
  basicstyle=\fontsize{7.2pt}{7.2pt}\ttfamily\selectfont,
  columns=fullflexible,
  breaklines=true,
  captionpos=b,
  commentstyle=\fontsize{7.2pt}{7.2pt}\color{col},
  keywordstyle=\fontsize{7.2pt}{7.2pt},
}

\begin{lstlisting}[language=python]
# f_q: encoder network for query
# f_k: encoder network for key
# queue: key dictionary
# r: momentum coefficient

f_k.param = f_q.param  # initialize parameters

# load a mini-batch x with N samples
for x in loader:
    nn = neighbor(x)  # generate neighbors of x
    x_q = augment(x)  # query randomly augmentation
    x_k = augment(x)  # key random augmentation
    x_n = augment(nn)  # neighbors random augmentation
    # randomly substitute half samples in x_k by x_n
    x_k = substitute(x_k, x_n)
    
    q = f_q.forward(x_q)  # queries: NxC
    k = f_k.forward(x_k)  # keys: NxC
    k = k.detach()  # no gradient to keys

    # positive logits: Nx1
    # bmm: batch matrix multiplication
    l_pos = bmm(q.view(N,1,C), k.view(N,C,1))
    # negative logits: NxK
    # mm: matrix multiplication
    l_neg = mm(q.view(N,C), queue.view(C,K))

    # logits: Nx(1+K)
    logits = cat([l_pos, l_neg], dim=1)

    # contrastive loss computation
    labels = zeros(N)
    loss = CrossEntropyLoss(logits/t, labels)

    # back propagation, only update the query network
    loss.backward()
    update(f_q.param)

    # dictionary update
    f_k.param = r*f_k.param+(1-r)*f_q.param
    enqueue(queue, k)  # push the current key
    dequeue(queue)  # pop the earliest key
\end{lstlisting}
\end{algorithm}

\textbf{Nearest Neighbor-Based Contrastive Learning.} Algorithm \ref{alg_code} provides the pseudo code of the proposed nearest neighbor-based contrastive learning. In the proposed framework, a set of sample pairs are selected from HSI and LiDAR data centered at the same position. During training, each sample pair is handled by random data augmentation to generate a query sample $x^q$ and a key sample $x^k$. They are encoded to embeddings $q$ and $k$, respectively. The embeddings from the same image are defined as positive key, and embeddings from different images are defined as negative key. A large number of negative key embeddings are stored in a dictionary $\{k_1, k_2, k_3, \ldots \}$, while one positive key $k_+$ is stored separately. Furthermore, we randomly select some nearest neighbors of $q$ to generate embeddings, which are denoted as $k_{n+}$. Next, in a minibatch, half positive keys $k_+$ are substituted by $k_{n+}$ to form new positive keys. Hence, nearest neighbors act small semantic perturbations. In our implementations, nearest neighbors denotes a region whose overlap area with $x^q$ is greater than 80\%.

We calculate the cosine similarities between $q$ and keys (both the positive key and negative keys). Then, the results are stored as $\{\mathcal{D}_+,\mathcal{D}_1, \mathcal{D}_2, \mathcal{D}_3, \ldots, \mathcal{D}_K\}$. Here $\mathcal{D}_+$ is the similarity between $q$ and positive key $k_+$. The rest are the similarities between $q$ and negative keys. $K$ is the number of negative keys.

The objective of contrastive learning is to force the query to match the positive key and far apart from the negative keys. To be specific, the contrastive loss whose value is low when $q$ is similar to the positive key $k_+$ and dissimilar to all the negative keys. Therefore, the contrastive loss function is designed as follows:
\begin{equation}
    \mathcal {L}=-\log{\frac{\exp(\mathcal{D}_+ / \tau)}
    {\sum^K_{i=1}{\exp( \mathcal{D}_i/\tau)}}}
\end{equation}
where $\tau$ is a temperature hyperparameter. Intuitively, softmax classifier and cross-entropy loss can be combined into the above equation.

\textbf{Dictionary Update and Moving Average Encoder.} Similar to MoCo \cite{he20cvpr}, we maintain the dictionary as a queue which stores many minibatch of negative samples. The negative samples in the dictionary are updated progressively. Specifically, during training, when a new minibatch is pushed into the dictionary, the oldest minibatch is removed. The length of the dictionary is flexibly set as a hyperparameter. 

Furthermore, the parameters of the encoder for the dictionary are updated slowly. Similar to MoCo, we use a separate moving average encoder for the key samples. During training, no backpropagation is done for the key encoder. The parameters of key encoder are updated as follows:
\begin{equation}
    \theta_k= r\theta_k +(1-r)\theta_q,
\end{equation}
where $\theta_k$ denotes the parameters of the key encoder, and $\theta_q$ denotes the parameters of the query encoder. $r$ is a momentum coefficient that controls the speed of key encoder update. Only $\theta_q$ is updated by backpropagation during training. In our implementations, $r$ is set to 0.9, since a slowly evolving key encoder is critical for robust feature learning.

\textbf{Shuffling BN.} Batch Normalization (BN) is employed in the encoder to speed up convergence and improve the generalization of the network. Similar to MoCo, we use the shuffling BN for better feature representation. In particular, we shuffle the sample order in the current minibatch for the key encoder. The sample order of the mini-batch for the query encoder is not changed.

\subsection{Bilinear Attention-Based Multisource Encoder}

\begin{figure}[]
    \centering
    \includegraphics[width=3.0in]{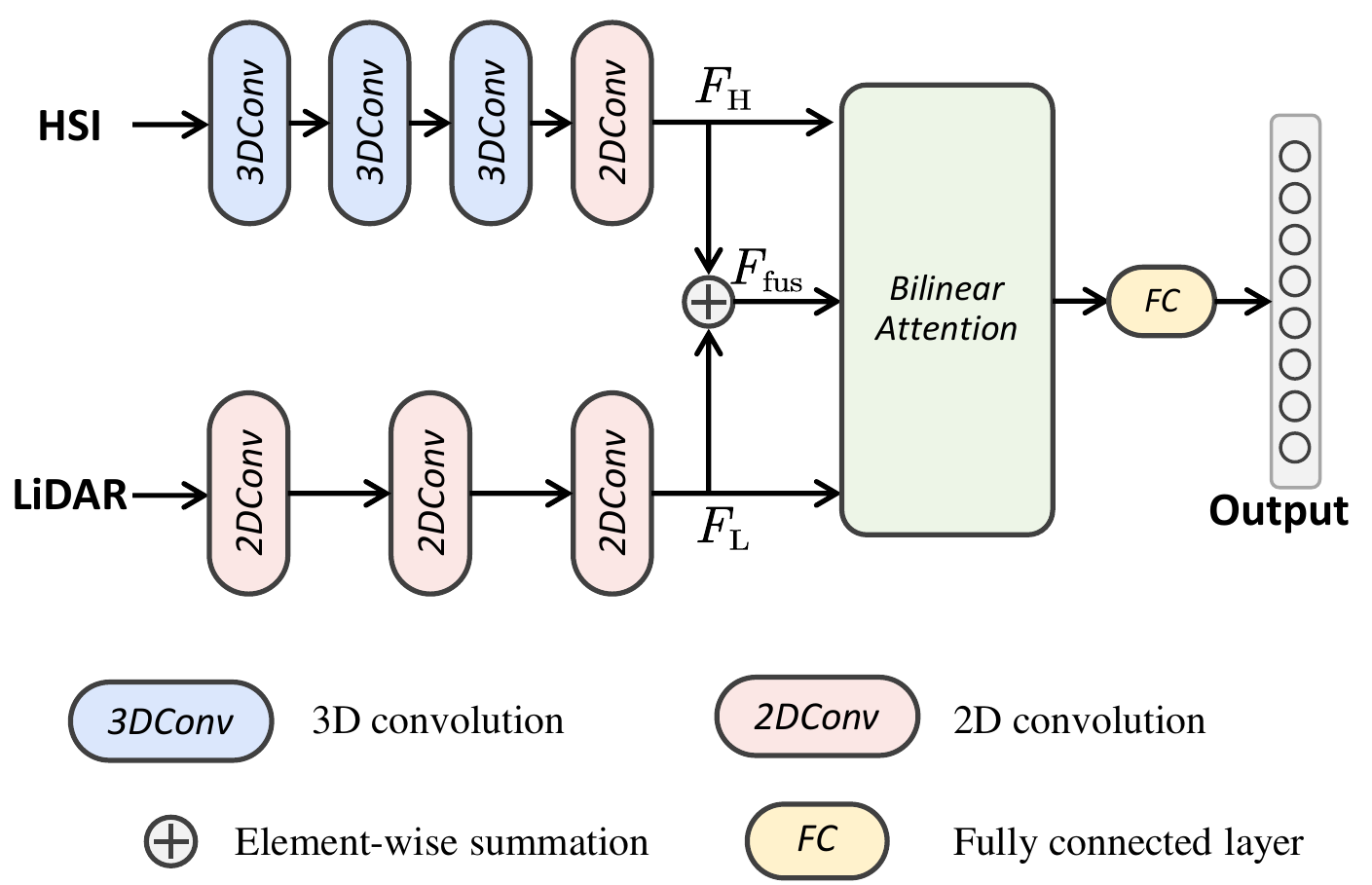}
    \caption{Bilinear attention-based multisource encoder.}
    \label{fig_encoder}
\end{figure}

In this work, the purpose of contrastive learning is to generate a pretrained model, and the model can be used for the classification task. To achieve high classification accuracy, the encoder is an essential part of the contrastive framework. We design a multisource encoder for hyperspectral and LiDAR feature modeling, as illustrated in Fig. \ref{fig_encoder}. It contains three parts: HSI feature extraction, LiDAR feature extraction, and bilinear attention fusion.

The detailed summary of the encoder in terms of layer type, kernel size, and output size is illustrated in Table \ref{table_encoder}. Hyperspectral data adopt a network similar to HybridSN \cite{roy2019hybridsn}, which uses both 3D and 2D convolutions for feature extraction. Three 3D convolution layers and one 2D convolution layer are used to derive the HSI feature $F_\textrm{H}$. At the same time, three 2D convolution layers are used to generate the LiDAR feature $F_\textrm{L}$. Next, $F_\textrm{H}$ and $F_\textrm{L}$ are combined to form the fused feature. A 2D convolution is used for feature embedding. Then, the fused feature $F_\textrm{fus}$ has the same dimension as $F_\textrm{H}$ and $F_\textrm{L}$. To effectively reduce the inherent redundancy in HSI, and thereby reduce the amount of data that needs to be processed in classification, Principal Component Analysis (PCA) is used to select the best 30 spectral bands for HSI feature extraction.

Finally, $F_\textrm{H}$, $F_\textrm{L}$ and $F_\textrm{fus}$ are fed into the bilinear attention fusion module as $\bf{Q}$, $\bf{K}$, and $\bf{V}$, respectively. The output of the bilinear attention fusion module is fed into a fully connected layer to generate the final feature for classification.

\begin{table}[]
\centering
\caption{Summary of the Proposed Multi-Source Encoder}
\renewcommand\arraystretch{1.3}
\begin{tabular}{cccc}
\toprule
\multicolumn{4}{c}{HSI feature extraction subnetwork} \\
\midrule
~ \# ~ & Layer type & Kernel number@size & Output size\\
\midrule
 &  Input & --- & (11, 11, 30, 1) \\
1 & 3D Conv & 8@3$\times$3$\times$9 & (9, 9, 22, 8) \\
2 & 3D Conv & 16@3$\times$3$\times$7 & (7, 7, 16, 16) \\
3 & 3D Conv & 32@3$\times$3$\times$5 & (5, 5, 12, 32) \\
4 & Reshape & --- & (5, 5, 384) \\
5 & 2D Conv & 256@3$\times$3 & (5, 5, 256) \\
6 & Reshape & --- & (25, 256) \\
\bottomrule
\toprule
\multicolumn{4}{c}{LiDAR feature extraction subnetwork} \\
\midrule
\# & Layer type & Kernel numbre@size & Output size \\
\midrule
& Input &  --- & (11, 11, 1) \\
1 & 2D Conv & 64@3$\times$3 & (9, 9, 64) \\
2 & 2D Conv & 128@3$\times$3 & (7, 7, 128) \\
3 & 2D Conv & 256@3$\times$3 & (5, 5, 256) \\
4 & Reshape & --- & (25, 256) \\
\bottomrule
\end{tabular}
\label{table_encoder}
\end{table}

\begin{figure}[]
    \centering
    \includegraphics[width=3.5in]{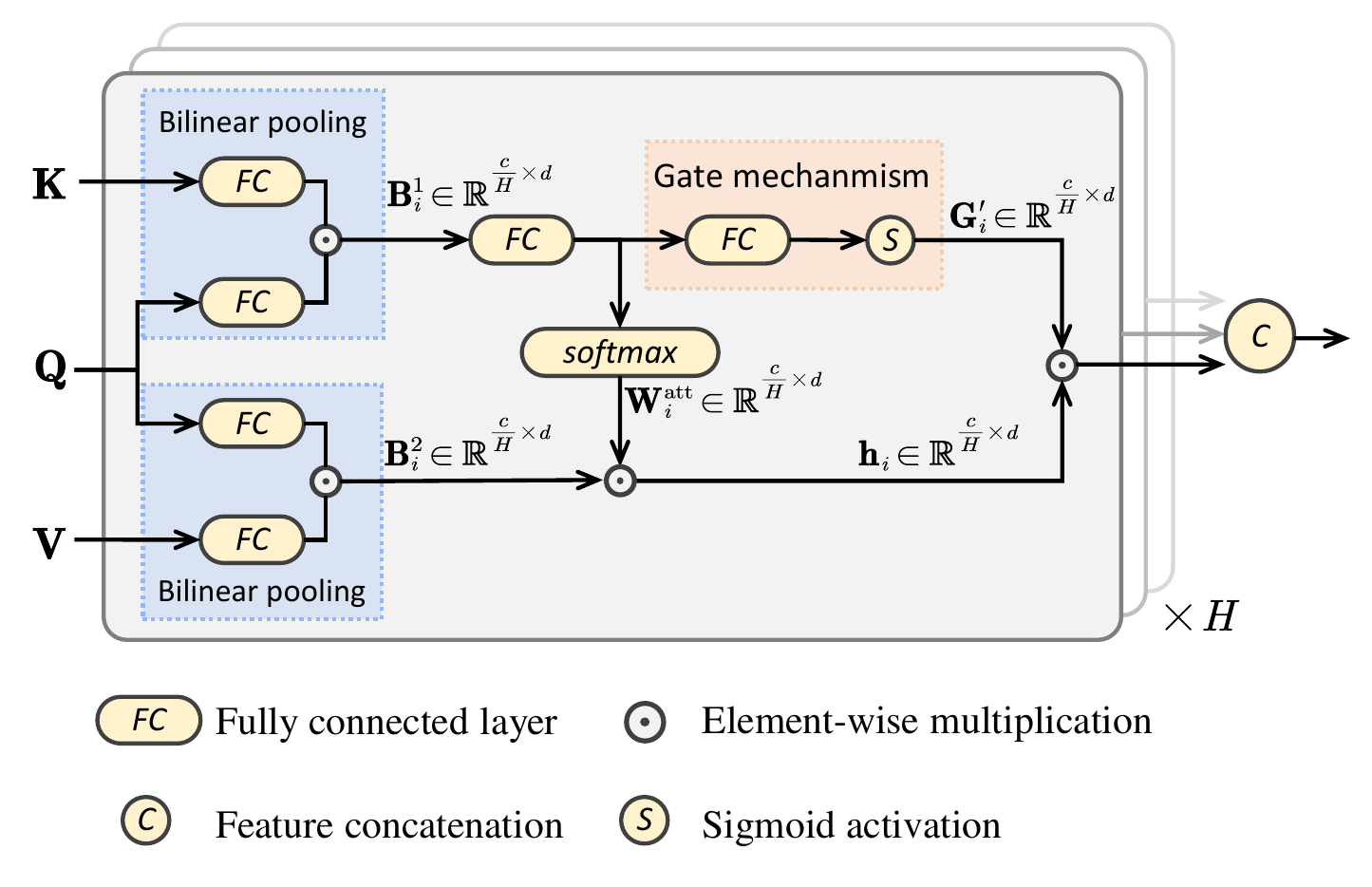}
    \caption{Bilinear attention fusion module. It can capture second-order interactions between multisource data.}
    \label{fig_bam}
\end{figure}

\subsection{Bilinear Attention Fusion Module}

The attention mechanism has made valuable breakthroughs in deep neural networks and has been successfully applied to cross-modal tasks (e.g., visual question answering \cite{yu19cvpr}, image captioning \cite{yan22tcsvt}, and image-text matching \cite{xu20tnnls}). This prompts recent methods to adopt the attention to trigger the interaction between multi-modal remote sensing data \cite{zheng22tgrs} \cite{liu22grsl} \cite{zhuang22grsl} \cite{zhang22tgrs}. In the conventional attention mechanism, the attention weights are estimated via linearly fusing the inputs. However, we argue that conventional attention exploits the first-order feature interaction and is limited in complex multisource feature reasoning.

Toward this end, we propose a bilinear attention fusion module to exploit the second-order feature interactions between the hyperspectral and LiDAR data. As illustrated in Fig. \ref{fig_bam}, it mainly contains two parts: the multi-head bilinear attention and the gate mechanism.

\textbf{Multi-Head Bilinear Attention.} Suppose we have query ${\bf{Q}}\in\mathbb{R}^{c\times d}$, key ${\bf{K}}\in\mathbb{R}^{c\times d}$, and value ${\bf{V}}\in\mathbb{R}^{c\times d}$, where $d$ denotes the feature dimension, and $c$ is the number of channels. To enhance the capability of feature representation, the multi-head scheme is used to model feature interactions from different subspaces as:
\begin{equation}
    {\bf{h}}_i=\textrm{BiAttention}({\bf{Q}}_i, {\bf{K}}_i, {\bf{V}}_i),
\end{equation}
where ${\bf{h}}_i$ is the output of the $i$-th head, and BiAttention denotes the bilinear attention. The number of heads is denoted by $H$. 

The bilinear attention first maps ${\bf{Q}}_i\in\mathbb{R}^{\frac{c}{H}\times d}$ and ${\bf{K}}_i\in\mathbb{R}^{\frac{c}{H}\times d}$ into a joint space as:
\begin{equation}
    {\bf{B}}^1_i=\sigma({\bf{Q}}_i{\bf{W}}^1_q)\odot \sigma({\bf{K}}_i{\bf{W}}_k),
\end{equation}
where ${\bf{W}}^1_q\in\mathbb{R}^{d\times d}$ and ${\bf{W}}_k\in\mathbb{R}^{d\times d}$ are weighting matrices, $\sigma$ is the ReLU activation and $\odot$ denotes the element-wise multiplication. As such, ${\bf{B}}^1_i\in \mathbb{R}^{\frac{c}{H}\times d}$ denotes the second-order representation between query ${\bf{Q}}_i$ and key ${\bf{K}}_i$.

Similarly, we compute the bilinear representation between ${\bf{Q}}_i$ and ${\bf{V}}_i$ as:
\begin{equation}
    {\bf{B}}^2_i=\sigma({\bf{Q}}_i{\bf{W}}^2_q)\odot \sigma({\bf{V}}_i{\bf{W}}_v),
\end{equation}
where ${\bf{W}}^2_q\in\mathbb{R}^{d\times d}$ and ${\bf{W}}_v\in\mathbb{R}^{d\times d}$ are weighting matrices. ${\bf{B}}^2_i\in \mathbb{R}^{\frac{c}{H}\times d}$ denotes the second-order representation between query ${\bf{Q}}_i$ and value ${\bf{V}}_i$.

Next, the bilinear representation ${\bf{B}}^1_i$ is projected into attention weights ${\bf{W}}^\textrm{att}_i\in\mathbb{R}^{\frac{c}{H}\times d}$ via a linear layer and a softmax layer as follows:
\begin{equation}
\hat{\bf{B}}^{1}_i=\sigma({{\bf{W}}_B\bf{B}}^1_i),
\end{equation}
\begin{equation}
    {\bf{W}}^\textrm{att}_i = \textrm{softmax}(\hat{\bf{B}}^{1}_i),
\end{equation}
where ${\bf{W}}_B\in\mathbb{R}^{\frac{c}{H}\times \frac{c}{H}}$ is the weight matrix. Next, the attended feature ${\bf{h}}_i\in\mathbb{R}^{\frac{c}{H}\times d}$ is derived by enhancing the attention weights as:
\begin{equation}
    {\bf{h}}_i= {\bf{W}}^{\textrm{att}}_i\odot {\bf{B}}^2_i
\end{equation}

\textbf{Gate Mechanism.} The aforementioned bilinear attention exploits the feature interactions among ${\bf{Q}}_i$, ${\bf{K}}_i$, and ${\bf{V}}_i$. However, there may contain noisy information in the query and key. To adaptively enhance the informative parts and suppress the useless parts, we design a gate mechanism. To be specific, for the $i$-th head, $\hat{\bf{B}}^{1}_i$ is fed into a linear layer and then handled with a sigmoid function to compute a weight mask ${\bf{G}}_i\in\mathbb{R}^{\frac{c}{H}\times1}$ as:
\begin{equation}
    {\bf{G}}_i = \textrm{sigmoid}(\hat{\bf{B}}^{1}_i {\bf{W}}_{B'}),
\end{equation}
where ${\bf{W}}_{B'}\in\mathbb{R}^{d\times 1}$ is the weight matrix. Next, ${\bf{G}}_i$ is expanded to form ${\bf{G}}'_i\in\mathbb{R}^{\frac{c}{H}\times d}$. Then the obtained gating mask is applied to control the information flow of ${\bf{h}}_i\in\mathbb{R}^{\frac{c}{H}\times d}$ as:
\begin{equation}
    \hat{\bf{h}}_i={\bf{G}}'_i\odot{\bf{h}}_i.
\end{equation}

Finally, by concatenating the results of multiple heads, we obtain the fused representation of multi-source data. In this work, the size of $\bf{Q}$, $\bf{K}$ and $\bf{V}$ is 25$\times$256. The number of heads $H$ is set to 5.

\section{Experimental Results and Analysis}

To validate the effectiveness of the proposed NNCNet, we conduct extensive experiments on four widely used benchmark datasets: Houston 2013 dataset, Trento dataset, MUUFL dataset and Houston 2018 dataset. We first compare the proposed NNCNet with state-of-the-art methods. Then we implemented additional evaluations to investigate the effectiveness of each component of our method.

\subsection{Datasets and Evaluation Metric}

\begin{table}
\centering
\caption{Train-Test Distribution of Samples for the Houston 2013 Dataset.}
\renewcommand\arraystretch{1.3}
\begin{tabular}{cccc}
\toprule
 ~~ No. ~~ & ~~ Class Name ~~ & ~~ Training ~~ & ~~ Test ~~\\
\midrule
 1 & Healthy grass & 198 & 1053 \\
 2 & Stressed grass & 190 & 1064 \\
 3 & Synthetic grass & 192 & 505 \\
 4 & Tree & 188 & 1056 \\
 5 & Soil & 186 & 1056 \\
 6 & Water & 182 & 143 \\
 7 & Residential & 196 & 1072 \\
 8 & Commercial & 191 & 1053 \\
 9 & Road & 193 & 1059 \\
 10 & Highway & 191 & 1036 \\
 11 & Railway & 181 & 1054 \\
 12 & Parking lot 1 & 192 & 1041 \\
 13 & Parking lot 2 & 184 & 285 \\
 14 & Tennis court & 181 & 247 \\
 15 & Running track & 187 & 473 \\
\midrule
 & Total & 2832 & 12197 \\
\bottomrule
\end{tabular}
\label{table1}
\end{table}

\begin{table}
\centering
\caption{Train-Test Distribution of Samples for the Trento Dataset.}
\renewcommand\arraystretch{1.3}
\begin{tabular}{cccc}
\toprule
~~ No. ~~ & ~~ Class Name ~~ & ~~ Training ~~ & ~~ Test ~~\\
\midrule
 1 & Apple trees & 129 & 3905 \\
 2 & Buildings & 125 & 2778 \\
 3 & Ground & 105 & 374 \\
 4 & Wood & 154 & 8969 \\
 5 & Vineyard & 184 & 10317 \\
 6 & Roads & 122 & 3052 \\
\midrule
   & Total & 819 & 29595 \\
\bottomrule
\end{tabular}
\label{table2}
\end{table}

\begin{table}
\centering
\caption{Train-Test Distribution of Samples for the MUUFL Dataset.}
\renewcommand\arraystretch{1.3}
\begin{tabular}{cccc}
\toprule
~~ No. ~~ & ~~ Class Name ~~ & ~~ Training ~~ & ~~ Test ~~\\
\midrule
1 & Trees & 150 & 23096 \\
2 & Mostly grass & 150 & 4120 \\
3 & Mixed ground surface & 150 & 6732 \\
4 & Dirt and sand & 150 & 1676 \\
5 & Road & 150 & 6537 \\
6 & Water & 150 & 316 \\
7 & Building shadow & 150 & 2083 \\
8 & Building & 150 & 6090 \\
9 & Sidewalk & 150 & 1235 \\
10 & Yellow curb & 150 & 33 \\
11 & Cloth panels & 150 & 119 \\
\midrule
  & Total & 1650 & 52037 \\
\bottomrule
\end{tabular}
\label{table3}
\end{table}

\begin{table}
\centering
\caption{Train-Test Distribution of Samples for the Houston 2018 Dataset.}
\renewcommand\arraystretch{1.3}
\begin{tabular}{cccc}
\toprule
 ~~ No. ~~ & ~~ Class Name ~~ & ~~ Training ~~ & ~~ Test ~~\\
\midrule
 1 & Healthy grass & 500 & 9299 \\
 2 & Stressed grass & 500 & 32002 \\
 3 & Artificial turf & 68 & 616 \\
 4 & Evergreen trees & 500 & 13095 \\
 5 & Deciduous trees & 500 & 4521 \\
 6 & Bare earth  & 451 & 4065 \\
 7 & Water & 26 & 240 \\
 8 & Residential buildings  & 500 & 39272 \\
 9 & Non-residential buildings & 500 & 223252 \\
 10 & Roads & 500 & 45366 \\
 11 & Sidewalks  & 500 & 33529 \\
 12 & Crosswalks  & 151 & 1367 \\
 13 & Major thoroughfares & 500 & 45848 \\
 14 & Highways & 500 & 9365 \\
 15 & Railways  & 500 & 6437 \\
 16 & Paved parking lots & 500 & 11000 \\
 17 & Unpaved parking lots & 14 & 132 \\
 18 & Cars & 500 & 6047 \\
 19 & Trains  & 500 & 4869 \\
 20 & Stadium seats  & 500 & 6324 \\
\midrule
 & Total & 8210 & 496646 \\
\bottomrule
\end{tabular}
\label{table4}
\end{table}

\textbf{Houston 2013 dataset}: The dataset was captured by the National Airborne Center for Laser Mapping, and it was used as a challenge in the 2013 GRSS Data Fusion Contest. The HSI was captured by the CASI sensor (144 spectral bands at a resolution of 2.5 m). Coregistered LiDAR data with the same resolution are available. A total of 15029 ground truth samples are distributed in 15 classes. They are divided into train and test sets containing 2832 and 12197 pixels, respectively. We used standard training and test sets, and Table \ref{table1} lists the number of training and test samples.

\begin{table*}
\centering
\caption{Classification Accuracy (\%) on the Houston 2013 Dataset} \label{table_houston2013}
\renewcommand\arraystretch{1.3}
\begin{tabular}{c|ccccccc}
\toprule
Class & FusAtNet \cite{fusatnet} & TBCNN \cite{xu18tgrs} & EndNet \cite{endnet} & MDL \cite{hong21tgrs} & CCNN \cite{hang20tgrs} & S2ENet \cite{s2enet} & \cellcolor{bg}NNCNet (ours)\\
\midrule
Healthy grass      & 79.20 &  81.01 & 78.54 & 83.00 & \textbf{91.55} & 82.72 & \cellcolor{bg}81.84\\

Stressed grass     & 96.71 &  97.93 & 96.33 & 98.68 & \textbf{99.72} & 100.0 & \cellcolor{bg}\textbf{99.72}\\

Synthetic grass    & 97.82 &  99.60 & \textbf{100.0} & 99.80 & 99.60 & 99.60 & \cellcolor{bg}99.80\\

Tree               & 97.63 &  94.13 & 88.26 & 93.94 & 97.63 & 95.74 & \cellcolor{bg}\textbf{99.43} \\

Soil               & \textbf{100.0} & 98.86 & \textbf{100.0} & 99.05 & \textbf{100.0} & 99.81 & \cellcolor{bg}\textbf{100.0}\\

Water              & 91.61 &  97.90 & \textbf{100.0} & \textbf{100.0} & 95.80 & 97.20 & \cellcolor{bg}\textbf{100.0}\\

Residential        & 76.31 &  80.50 & 83.02 & 79.66 & 83.12 & 91.23 & \cellcolor{bg}\textbf{94.87}\\

Commercial         & 74.17 &  87.46 & 79.96 & 80.44 & 94.49 & 91.55 & \cellcolor{bg}\textbf{94.78}\\

Road               & 89.05 &  86.50 & 93.30 & 84.70 & 93.20 & 95.94 & \cellcolor{bg}\textbf{96.03}\\

Highway            & 92.86 &  64.86 & 92.28 & 94.88 & 89.96 & 84.75 & \cellcolor{bg}\textbf{99.81}\\

Railway            & 94.21 &  93.74 & 85.86 & 85.67 & 96.39 & 94.31 & \cellcolor{bg}\textbf{99.34}\\

Parking lot 1      & 87.32 &  74.93 & \textbf{99.81} & 98.75 & 99.71 & 97.79 & \cellcolor{bg}\textbf{99.81}\\

Parking lot 2      & 84.21 &  85.96 & 83.16 & 82.46 & 89.82 & 89.47 & \cellcolor{bg}\textbf{90.88}\\

Tennis court       & \textbf{100.0} &  \textbf{100.0} & \textbf{100.0} & \textbf{100.0} & \textbf{100.0} & \textbf{100.0} & \cellcolor{bg}\textbf{100.0}\\

Running track      & \textbf{100.0} &  \textbf{100.0} & \textbf{100.0} & \textbf{100.0} & \textbf{100.0} & \textbf{100.0} & \cellcolor{bg}\textbf{100.0}\\
\midrule
OA & 89.70 & 87.57 & 90.71 & 90.80 & 94.98 & 93.99 & \cellcolor{bg}\textbf{96.77}\\
AA & 90.73 & 89.55  & 92.03 & 92.06 & 95.40 & 94.67 & \cellcolor{bg}\textbf{97.06}\\
Kappa & 88.81 & 86.50  & 89.92 & 90.01 & 94.56 & 93.48 & \cellcolor{bg}\textbf{96.49}\\
\bottomrule
\end{tabular}
\end{table*}

\begin{table*}
\centering
\caption{Classification Accuracy (\%) on the Trento Dataset} \label{table_trento}
\renewcommand\arraystretch{1.3}
\begin{tabular}{c|ccccccc}
\toprule
Class & FusAtNet \cite{fusatnet} & TBCNN \cite{xu18tgrs} & EndNet \cite{endnet} & MDL \cite{hong21tgrs} & CCNN \cite{hang20tgrs} & S2ENet \cite{s2enet} & NNCNet (ours)\\
\midrule
Apple trees & 99.95 & 99.87 & \textbf{99.90} & \textbf{99.90} & \textbf{99.90} & \textbf{99.90} & \cellcolor{bg}99.13\\
Buildings   & 98.92 & 98.81 & 99.03 & \textbf{99.10} & \textbf{99.10} & 98.88 & \cellcolor{bg}98.92\\
Ground      & 85.56 & 81.02 & 85.83 & 86.36 & 86.90 & 86.36 & \cellcolor{bg}\textbf{99.73}\\
Wood        & \textbf{100.0} & \textbf{100.0} & \textbf{100.0} & \textbf{100.0} & \textbf{100.0} & \textbf{100.0} & \cellcolor{bg}\textbf{100.0}\\
Vineyard    & 99.68 & 98.40 & 99.31 & 99.61 & 99.67 & 99.21 & \cellcolor{bg}\textbf{100.0}\\
Roads       & \textbf{92.07} & 89.35 & 90.83 & 91.12 & 91.25 & 91.32 &\cellcolor{bg}91.88\\
\midrule
OA          & 98.77 & 97.96 & 98.52 & 98.66 & 98.71 & 98.53 & \cellcolor{bg}\textbf{98.92}\\
AA          & 96.03 & 94.57 & 95.81 & 96.01 & 96.13 & 95.94 & \cellcolor{bg}\textbf{98.26}\\
Kappa       & 98.35 & 97.27 & 98.01 & 98.21 & 98.27 & 98.03 & \cellcolor{bg}\textbf{98.55}\\
\bottomrule
\end{tabular}
\end{table*}

\begin{figure*}
  \centering
  \includegraphics[width=0.95\textwidth]{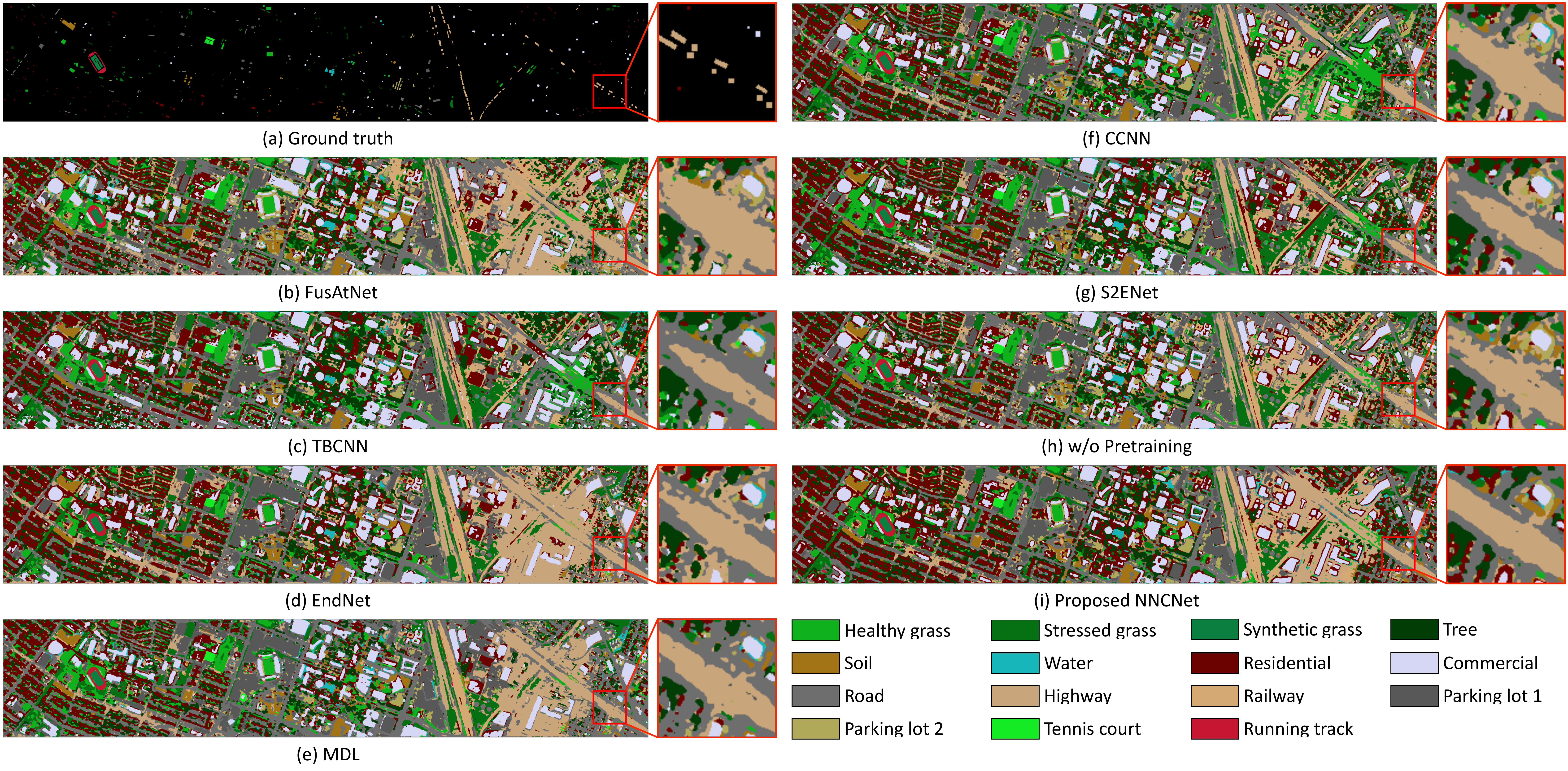}\\
  \caption{Classification maps for the Houston 2013 dataset. (a) Groundtruth. (b) FusAtNet. (c) TBCNN. (d) EndNet. (e) MDL. (f) CCNN. (g) S2ENet. (h) Proposed NNCNet without pretraining. (i) Proposed NNCNet.} \label{fig_houston2013_map}
\end{figure*}

\textbf{Trento dataset}: The dataset was collected in a rural region south of Trento, Italy. The HSI image consists of 63 bands with a wavelength range of 0.42-0.99 $\mu$m. The size of the datset is 166$\times$660 pixels, and the spatial resolution of the datset is 1.0 m. A total of 30214 ground truth samples are distributed in 6 classes. Table \ref{table2} lists the distribution of training and test samples for the Trento dataset.

\textbf{MUUFL dataset}: The MUUFL dataset is captured over the University of Southern Mississippi Gulf Coast campus in November 2010. The HSI contains 72 spectral bands, but the first and last four bands are removed for noise reduction, leaving 64 bands for classification. The total size of the dataset is 325$\times$220 pixels. Table \ref{table3} lists the training and test samples available for the dataset. In our experiments, we use the entire data in the pretraining phase, while in the training validation phase, we use only the portion of the training set for which labels are given.

\textbf{Houston 2018 dataset}: The dataset was captured by the Hyperspectral Image Analysis Laboratory and the National Center for Airborne Laser Mapping (NCALM) at the University of Houston. It was originally released for the 2018 IEEE GRSS Data Fusion Contest. Hyperspectral data covers 380-1050 nm spectral range with 48 bands at 1.0 m ground sample distance. The dataset contains a total of 4768$\times$1202 pixels in which a piece is delineated as the training set with the size of 2384$\times$601 pixels. Table \ref{table4} lists the distribution of training and test samples for the Houston 2018 dataset.

The performance of the model is evaluated by Overall Accuracy (OA), Average Accuracy (AA), and Kappa coefficient. OA denotes the ratio of the model's correct predictions to the overall number on all test samples. AA is the ratio between the number of correct predictions in each category and the overall number in each category, and finally the average of the accuracy in each category. Kappa is the percentage of agreement corrected by the number of agreements that would be expected purely by chance.

\subsection{Implementation Details}

The proposed contrastive learning architecture is used to generate a pretrained model. In the contrastive learning phase, we use the Adam optimizer. The mini-batch size is set to 64 and the learning rate is set to 0.0005. The image patch of 11$\times$11 pixels is randomly cropped from the dataset as training samples. 

After obtaining the pretrained model, training samples from the dataset are used for fine-tuning the model. In the fine-tuning phase, the mini-batch size is set to 128, and the setting of the optimizer is the same as that in the contrastive learning phase.

\subsection{Classification Accuracy and Discussion}

\begin{figure*}
  \centering
  \includegraphics[width=5.5in]{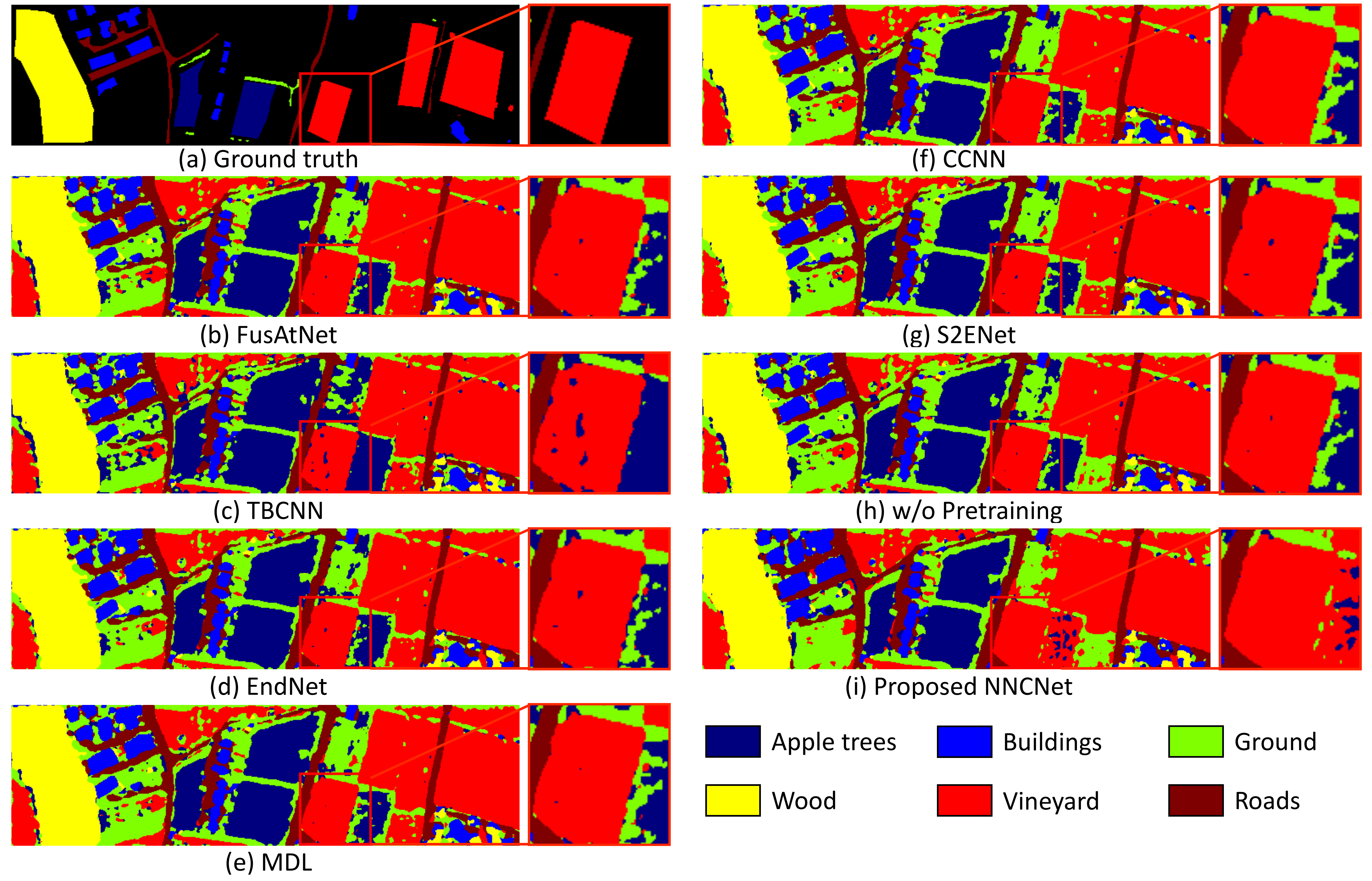}
  \caption{Classification maps for the Trento dataset. (a) Groundtruth. (b) FusAtNet. (c) TBCNN. (d) EndNet. (e) MDL. (f) CCNN. (g) S2ENet. (h) Proposed NNCNet without pretraining. (i) Proposed NNCNet.}\label{fig_trento_map}
\end{figure*}

\begin{table*}
\centering
\caption{Classification Accuracy (\%) on the MUUFL Dataset} \label{table_muufl}
\renewcommand\arraystretch{1.3}
\begin{tabular}{c|ccccccc}
\toprule
Class & FusAtNet \cite{fusatnet} & TBCNN \cite{xu18tgrs} & EndNet \cite{endnet} & MDL \cite{hong21tgrs} & CCNN \cite{hang20tgrs} & S2ENet \cite{s2enet} & NNCNet (ours)\\
\midrule
Trees                  & \textbf{95.31} & 91.18 & 90.86 & 90.95 & 92.40 & 93.91 & \cellcolor{bg}93.09\\
Mostly grass           & 79.83 & 83.98 & 83.30 & 84.54 & 83.52 & \textbf{88.28} & \cellcolor{bg}86.82\\
Mixed ground surface   & 83.69 & 83.72 & 84.27 & 83.01 & 84.34 & 81.85 & \cellcolor{bg}\textbf{86.29}\\
Dirt and sand          & \textbf{97.73} & 96.12 & 96.00 & 96.42 & 96.72 & 97.32 & \cellcolor{bg}96.18\\
Road                   & 84.86 & 91.23 & 91.11 & 90.44 & 91.68 & 91.28 & \cellcolor{bg}\textbf{92.35}\\
Water                  & \textbf{99.68} & \textbf{99.68} & \textbf{99.68} & \textbf{99.68} & \textbf{99.68} & \textbf{99.68} & \cellcolor{bg}\textbf{99.68}\\
Building shadow        & 83.01 & \textbf{92.85} & 92.61 & 92.75 & 92.61 & 88.29 & \cellcolor{bg}92.75\\
Building               & 94.70 & 96.86 & \textbf{96.90} & 96.70 & 96.80 & 95.99 & \cellcolor{bg}96.21\\
Sidewalk               & 89.80 & 87.85 & 88.34 & 87.85 & 89.23 & 88.50 & \cellcolor{bg}\textbf{91.34}\\
Yellow curb            & 87.88 & \textbf{90.91} & \textbf{90.91} & \textbf{90.91} & \textbf{90.91} & 87.88 & \cellcolor{bg}84.85\\
Cloth panels           & \textbf{99.16} & \textbf{99.16} & \textbf{99.16} & \textbf{99.16} & \textbf{99.16} & \textbf{99.16} & \cellcolor{bg}\textbf{99.16}\\
\midrule
OA                     & 90.68 & 90.53 & 90.39 & 90.27 & 91.20 & 91.61 & \cellcolor{bg}\textbf{92.07}\\
AA                     & 90.51 & 92.14 & 92.10 & 92.03 & 92.45 & 92.01 & \cellcolor{bg}\textbf{92.61}\\
Kappa                  & 87.65 & 87.59 & 87.42 & 87.26 & 88.44 & 88.93 & \cellcolor{bg}\textbf{89.56}\\
\bottomrule
\end{tabular}
\end{table*}

\begin{figure*}
  \centering
  \includegraphics[width=4.5in]{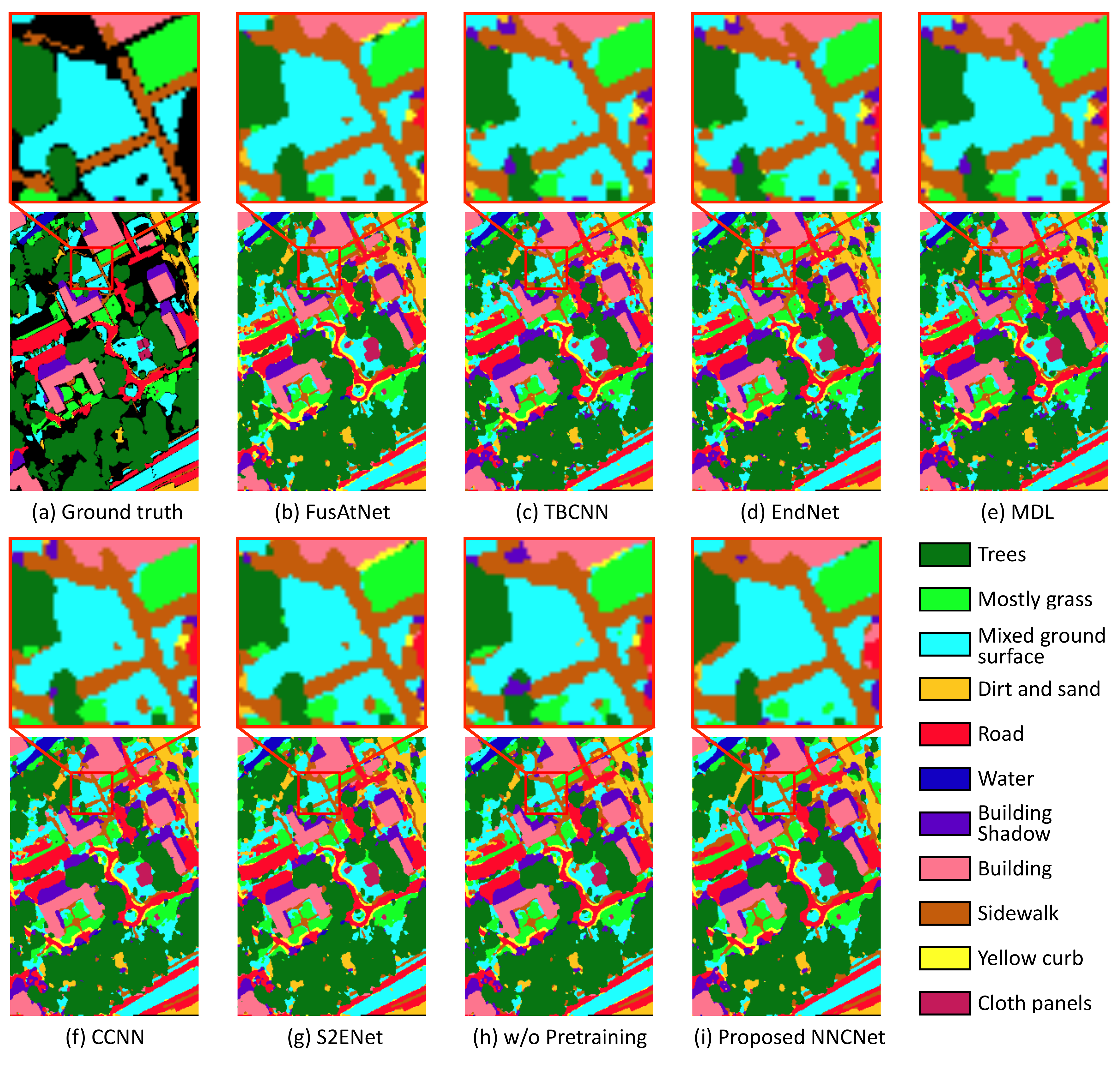}
  \caption{Classification maps for the MUUFL dataset. (a) Groundtruth. (b) FusAtNet. (c) TBCNN. (d) EndNet. (e) MDL. (f) CCNN. (g) S2ENet. (h) Proposed NNCNet without pretraining. (i) Proposed NNCNet.} \label{fig_muufl_map}
\end{figure*}

The proposed NNCNet is implemented on the Houston 2013, Trento, MUUFL and Houston 2018 datasets. To verify the effectiveness of the proposed NNCNet, we compared it with six state-of-the-art methods, including FusAtNet \cite{fusatnet}, TBCNN \cite{xu18tgrs}, EndNet \cite{endnet}, MDL \cite{hong21tgrs}, CCNN \cite{hang20tgrs}, and S2ENet \cite{s2enet}. In particular, FusAtNet \cite{fusatnet} exploits HSI and LiDAR features via cross-attention, and attentive spectral and spatial representations are combined to compute modality-specific feature embeddings. TBCNN \cite{xu18tgrs} uses a two-branch CNN for HSI and LiDAR feature extraction. In EndNet \cite{endnet}, a deep encoder–decoder network is utilized for multimodal information fusion and classification. MDL \cite{hong21tgrs} presents a general multimodal deep learning framework. CCNN \cite{hang20tgrs} presents a coupled network for multimodal information fusion. Feature-level and decision-level fusion are integrated for heterogeneous feature representation. S2ENet \cite{s2enet} is a spatial-spectral enhancement network that improves spatial and spectral feature representations simultaneously. For a fair comparison, all the compared methods adopt the default parameters provided in their works.

\begin{figure*}
  \centering
  \includegraphics[width=0.95\textwidth]{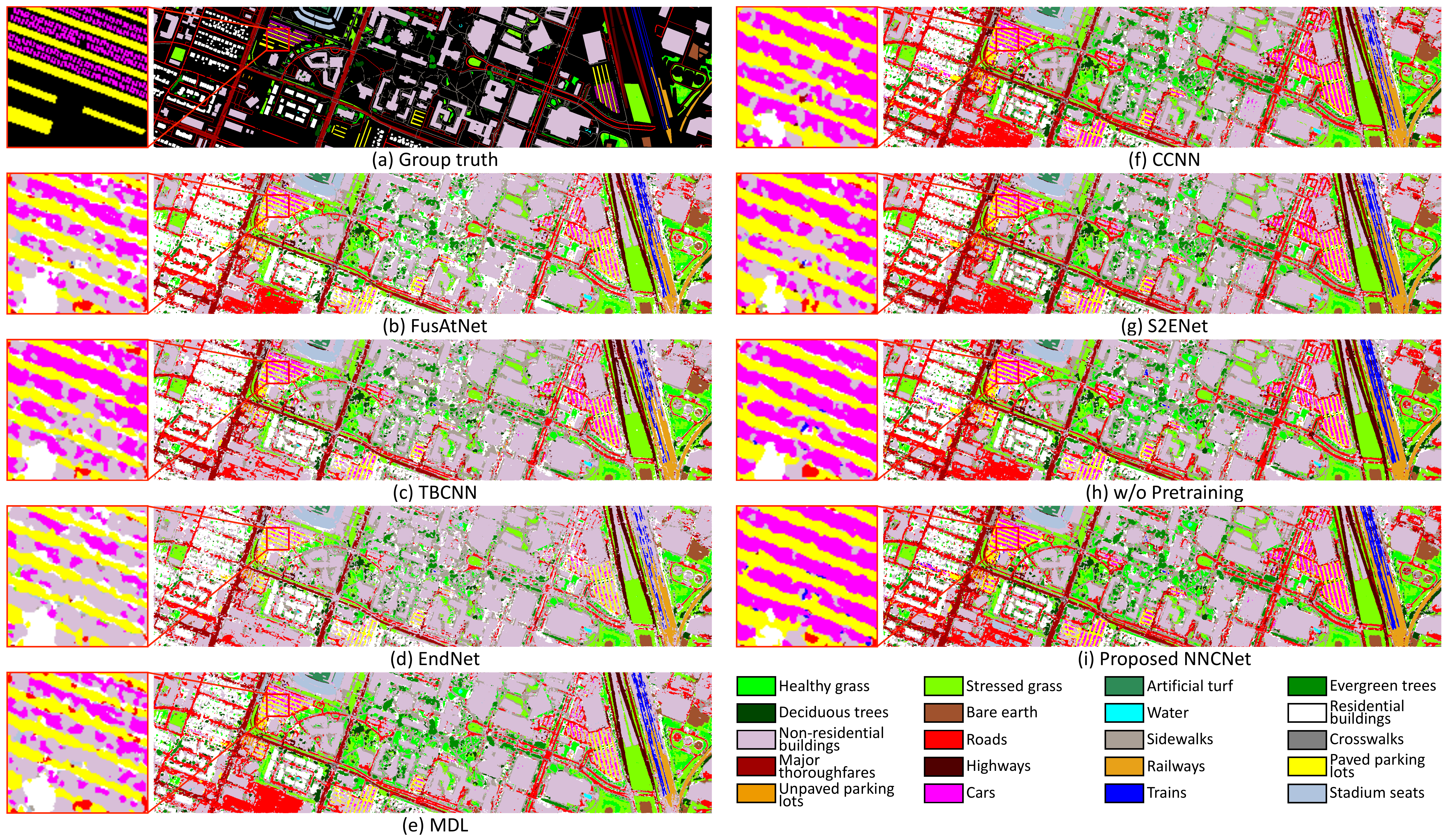}\\
  \caption{Classification maps for the Houston 2018 dataset. (a) Groundtruth. (b) FusAtNet. (c) TBCNN. (d) EndNet. (e) MDL. (f) CCNN. (g) S2ENet. (h) Proposed NNCNet without pretraining. (i) Proposed NNCNet.} \label{fig_houston2018_map}
\end{figure*}

\begin{table*}
\centering
\caption{Classification Accuracy (\%) on the Houston 2018 Dataset} \label{table_houston2018}
\renewcommand\arraystretch{1.3}
\begin{tabular}{c|ccccccc}
\toprule
Class & FusAtNet \cite{fusatnet} & TBCNN \cite{xu18tgrs} & EndNet \cite{endnet} & MDL \cite{hong21tgrs} & CCNN \cite{hang20tgrs} & S2ENet \cite{s2enet} & \cellcolor{bg}NNCNet (ours)\\
\midrule
Healthy grass             & 89.40 & 90.91 & 89.55 & \textbf{93.99} & 93.39 & 91.29 & \cellcolor{bg}93.36\\
Stressed grass            & 90.65 & 88.83 & 89.51 & 88.95 & 90.84 & \textbf{91.97} & \cellcolor{bg}91.95\\
Artificial turf           & 98.54 & 84.90 & 75.97 & \textbf{98.86} & 98.38 & 96.59 & \cellcolor{bg}98.38\\
Evergreen trees           & 85.30 & 71.97 & 67.97 & 90.60 & \textbf{94.25} & 88.91 & \cellcolor{bg}92.00\\
Deciduous trees           & 73.15 & 70.87 & 66.98 & 76.55 & \textbf{80.62} & 79.12 & \cellcolor{bg}76.86\\
Bare earth                & \textbf{100.0} & 99.78 & \textbf{100.0} & \textbf{100.0} & 99.48 & 99.78 & \cellcolor{bg}99.98\\
Water                     & \textbf{99.17} & 96.67 & 92.92 & 98.75 & 95.83 & 93.75 & \cellcolor{bg}96.25\\
Residential buildings     & \textbf{97.29} & 94.93 & 92.54 & 86.31 & 91.43 & 91.31 & \cellcolor{bg}87.58\\
Non-residential buildings & 94.36 & 95.78 & 96.78 & \textbf{97.75} & 93.93 & 95.22 & \cellcolor{bg}97.04\\
Roads                     & 62.29 & 53.26 & 42.71 & 69.65 & \textbf{73.14} & 70.95 & \cellcolor{bg}71.25\\
Sidewalks                 & 64.00 & 72.67 & 71.00 & 68.30 & \textbf{78.85} & 76.82 & \cellcolor{bg}70.63\\
Crosswalks                & 40.53 & 41.84 & 03.66 & 49.82 & 52.38 & \textbf{56.18} & \cellcolor{bg}38.33\\
Major thoroughfares       & 69.77 & 78.48 & 71.08 & 60.56 & 76.08 & 76.25 & \cellcolor{bg}\textbf{81.58}\\
Highways                  & 97.16 & 98.55 & 96.11 & 96.18 & \textbf{98.70} & 98.24 & \cellcolor{bg}98.54\\
Railways                  & 99.43 & 99.19 & 98.91 & 97.84 & 99.52 & 99.67 & \cellcolor{bg}\textbf{99.94}\\
Paved parking lots        & 85.68 & 78.73 & 75.27 & 82.50 & 87.32 & 91.12 & \cellcolor{bg}\textbf{95.25}\\
Unpaved parking lots      & \textbf{100.0} & \textbf{100.0} & \textbf{100.0} & \textbf{100.0} & \textbf{100.0} & \textbf{100.0} & \cellcolor{bg}\textbf{100.0}\\
Cars                      & 56.13 & 72.65 & 24.77 & 47.87 & 89.37 & 77.11 & \cellcolor{bg}\textbf{92.16}\\
Trains                    & 91.29 & 65.06 & 60.67 & 90.37 & 76.38 & 92.75 & \cellcolor{bg}\textbf{99.01}\\
Stadium seats             & 99.62 & 99.49 & 99.34 & 99.59 & \textbf{99.83} & 99.65 & \cellcolor{bg}99.78\\
\midrule
OA                        & 85.98 & 86.33 & 83.84 & 86.70 & 88.64 & 88.87 & \cellcolor{bg}\textbf{89.89}\\
AA                        & 84.68 & 82.72 & 75.78 & 84.72 & 88.48 & 88.33 & \cellcolor{bg}\textbf{88.99}\\
Kappa                     & 81.46 & 81.83 & 78.06 & 82.15 & 85.19 & 85.38 & \cellcolor{bg}\textbf{86.65}\\
\bottomrule
\end{tabular}
\end{table*}

\begin{figure*}[htb]
\centering
\includegraphics[width=1\textwidth]{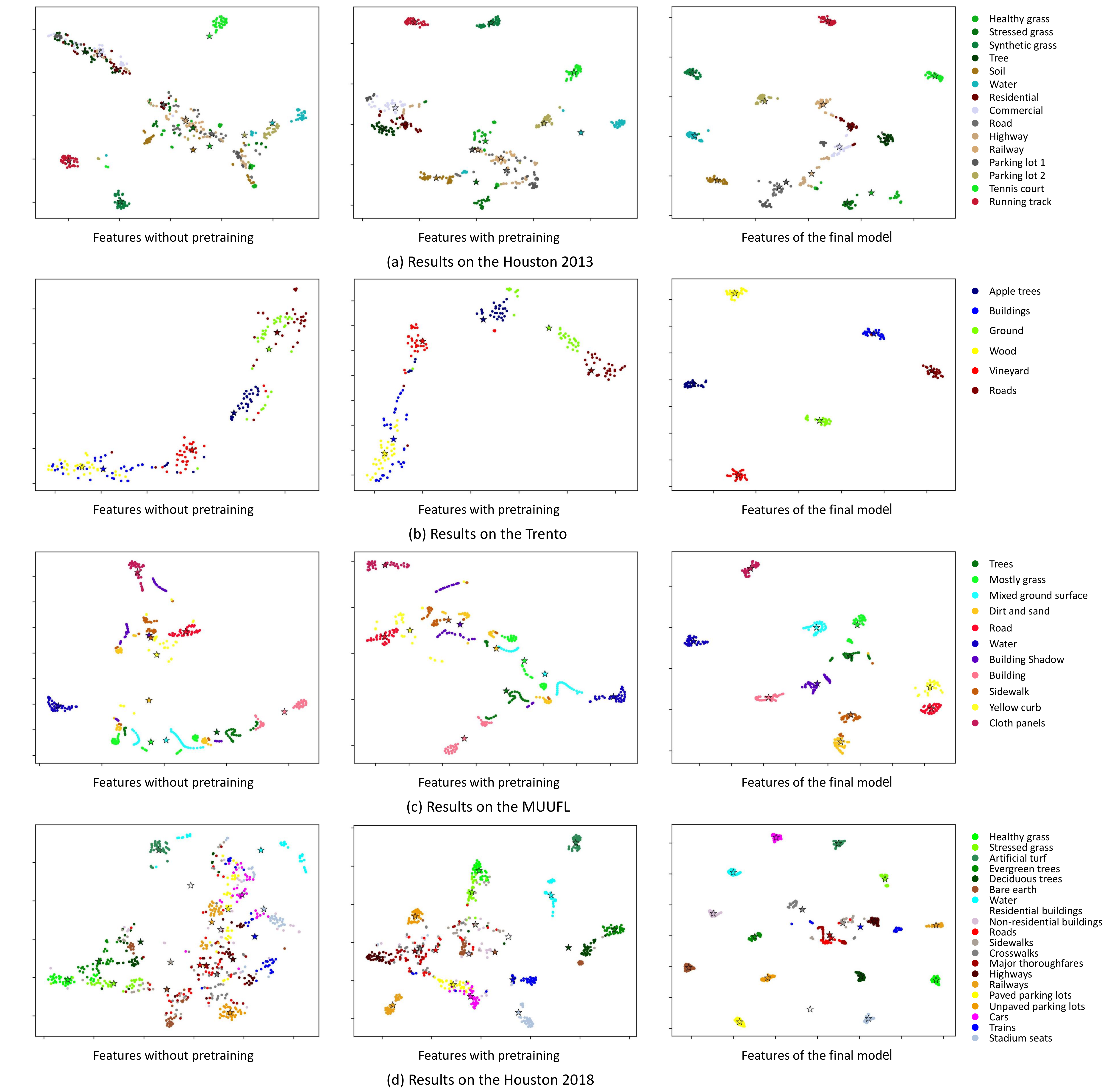}
\caption{Feature visualizations on different datasets. (a) Results on the Houston 2013 dataset. (b) Results on the Trento dataset. (c) Results on the MUUFL dataset. (d) Results on the Houston 2018 dataset. The first column denotes features without pretraining, the second column denotes features with pretraining, the last column represents the features of our final model. The star denotes the cluster center of each class of features.}
\label{fig_tsne}
\end{figure*}

Table \ref{table_houston2013} shows the classification results on the Houston 2013 dataset. The proposed NNCNet achieves the best performance in terms of OA, AA and Kappa coefficients. Our NNCNet outperforms the competitor (CCNN and S2ENet) by 1.78\% and 2.78\% for OA, respectively. It shows that the proposed self-supervised framework effectively models the correlations between multisource samples. Furthermore, the accuracy for `highway' class (99.81\%) is significantly improved by our NNCNet. There are many unlabeled highway regions in the Houston 2013 dataset. Therefore, our NNCNet captured the texture and spectral features of highway via contrastive learning from unlabeled data. The classification maps are illustrated in Fig. \ref{fig_houston2013_map}. It can be observed that without pretraining, some highway regions are falsely classified into road. In contrast, the proposed NNCNet performs better through contrastive learning.

\begin{table*}
\centering
\caption{Performance Comparison of Several Variants of the Proposed Model on Different Datasets} \label{table_ablation}
\renewcommand\arraystretch{1.3}
\begin{tabular}{c|cccc|cccc}
\toprule
Variant & Pretrain & \makecell{Bilinear\\Attention} & \makecell{Gate\\Mechanism} & \makecell{Nearest\\Neighbor} &  Houston 2013 & Trento & MUUFL & Houston 2018 \\
\midrule
1 & {\color{gray}\faTimes} & {\color{gray}\faTimes} & {\color{gray}\faTimes} & {\color{gray}\faTimes} & 95.20 & 98.74 & 91.38 & 88.21 \\
2 & \faCheck & {\color{gray}\faTimes} & {\color{gray}\faTimes} & {\color{gray}\faTimes} & 95.57 & 98.80 & 91.60 & 88.72 \\
3 & \faCheck & {\color{gray}\faTimes} & {\color{gray}\faTimes} & \faCheck & 96.30 & 98.88 & 91.83 & 89.41 \\
4 & \faCheck & \faCheck & {\color{gray}\faTimes} & {\color{gray}\faTimes} & 95.64 & 98.86 & 91.68 & 88.79 \\
5 & \faCheck & \faCheck & {\color{gray}\faTimes} & \faCheck & 96.47 & 98.90 & 92.01 & 89.76 \\
6 & \faCheck & \faCheck & \faCheck & {\color{gray}\faTimes} & 95.84 & 98.86 & 91.68 & 88.83 \\
\rowcolor{bg} 7 & \faCheck & \faCheck & \faCheck & \faCheck & 96.77 & 98.92 & 92.07 & 89.89 \\
\bottomrule
\end{tabular}
\end{table*}

\begin{figure*}[htbp]
\centering
\includegraphics[width=6in]{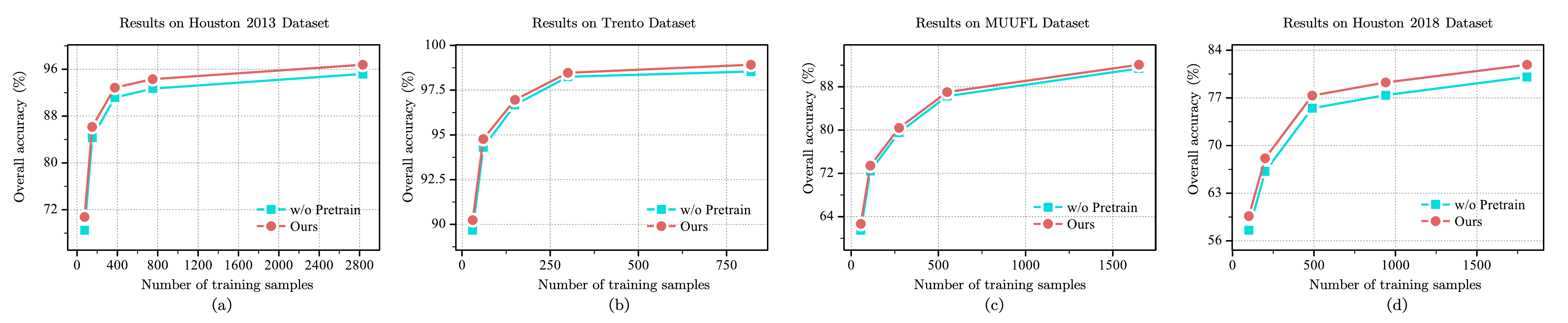}
\caption{Classification accuracy for different number of samples.}
\label{fig_sam_num}
\end{figure*}

\begin{figure}[htbp]
\centering
\includegraphics[width=3.4in]{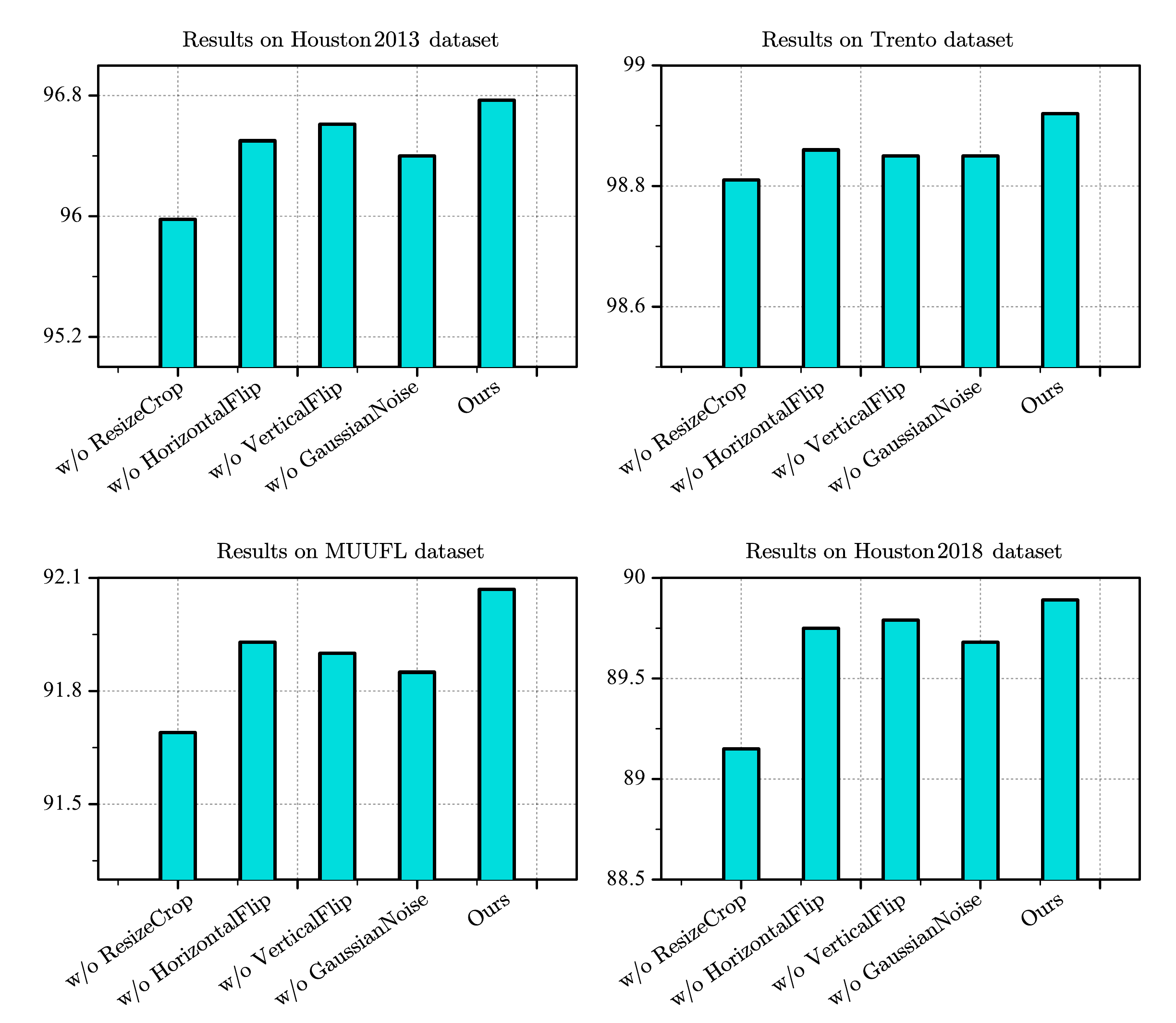}
\caption{Performance comparison of our model using different data augmentations.}
\label{fig_dataaug}
\end{figure}

\begin{figure*}[htbp]
\centering
\includegraphics[width=6in]{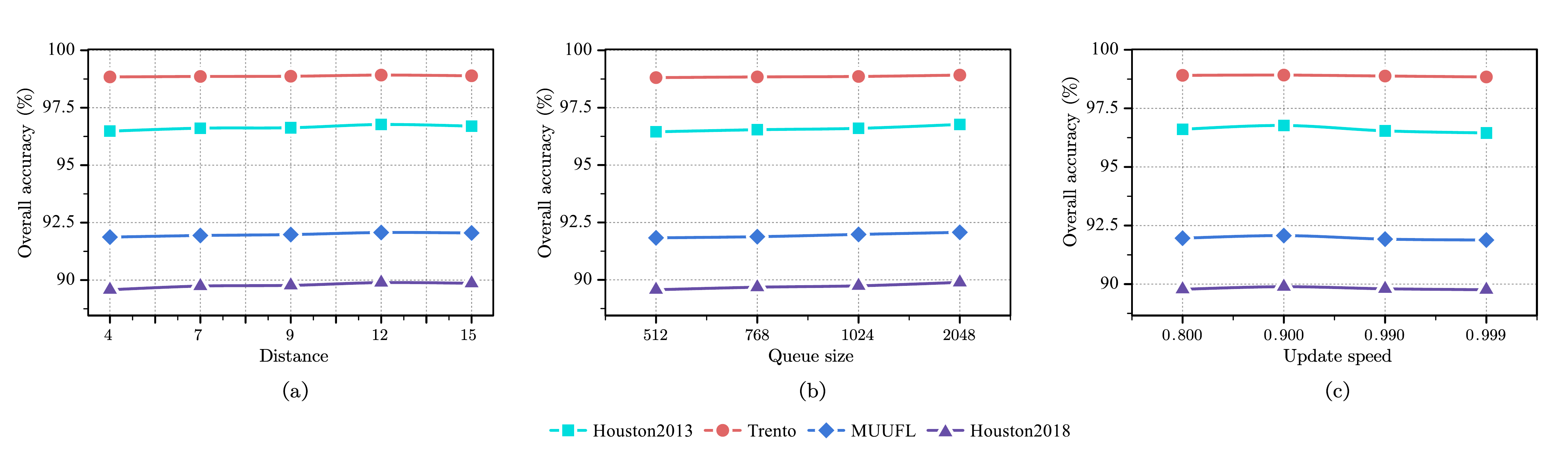}
\caption{Classification accuracy for different spatial distances, queue sizes, and mini-batch sizes on different datasets.}
\label{fig_queue_size}
\end{figure*}

\begin{figure}[htbp]
\centering
\includegraphics[width=2.6in]{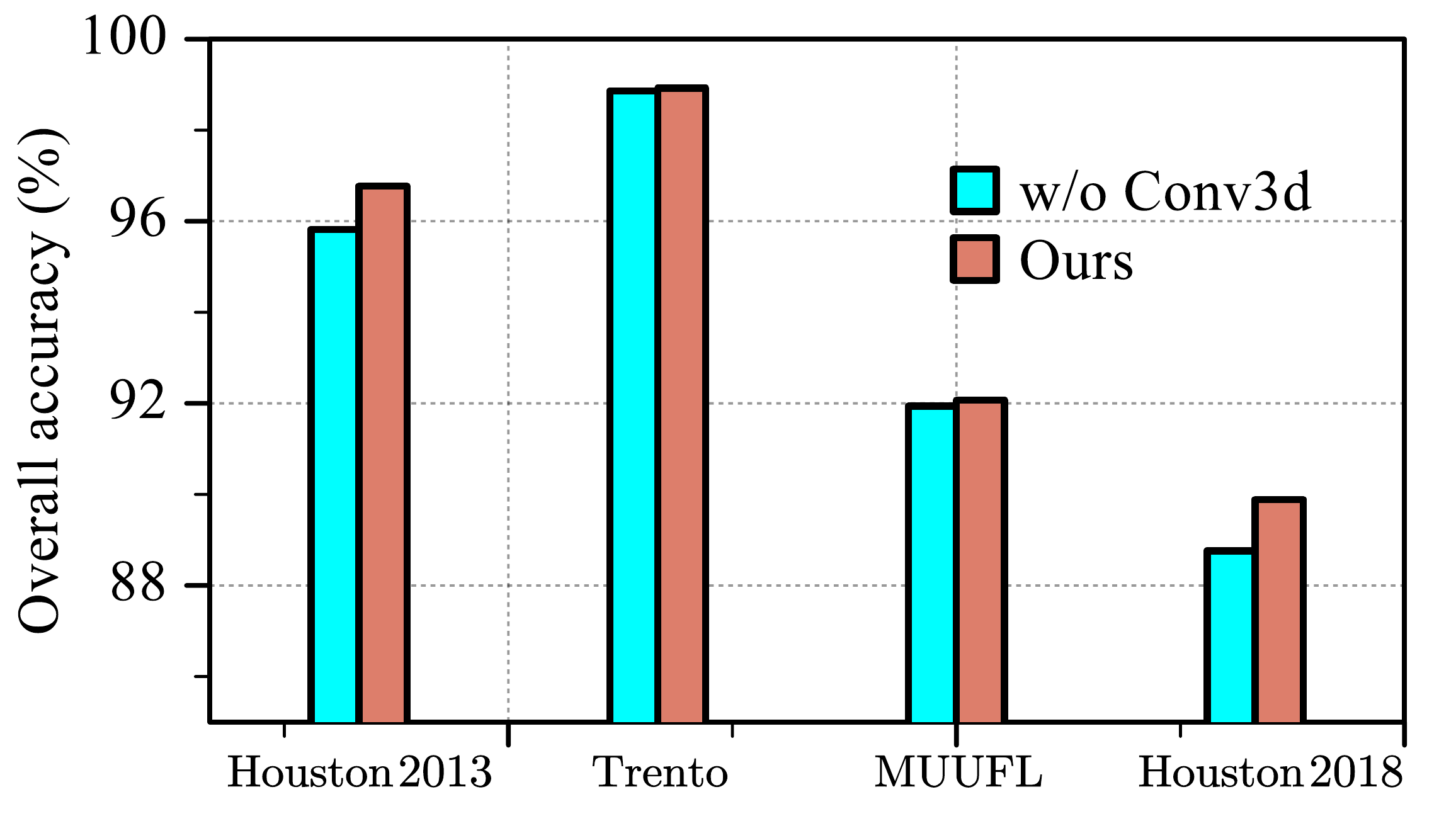}
\caption{Performance comparison of our model with or without 3D convolution.}
\label{fig_3dconv}
\end{figure}

Table \ref{table_trento} illustrates the classification results of different methods on the Trento dataset. The classification maps are shown in Fig. \ref{fig_trento_map}. It can be seen that without pretraining, some vineyard regions are falsely classified into apple trees. In addition, the proposed NNCNet achieves the best performance in terms of OA, AA, and Kappa. The proposed method achieves the best OA in `ground'. There is only a small amount of labeled data in this class, but it still accounts for a large portion of the entire graph. It is evident that our NNCNet is capable to learning the robust feature representations when training samples are limited.

Table \ref{table_muufl} shows the classification results of different methods on the MUUFL dataset. The proposed NNCNet obtains the best performance against the other methods. To be specific, the proposed method has the best OA (92.07\%) and reached the highest accuracy in five classes (Mixed ground surface, Road, Water, Sidewalk and Cloth panels). The classification results of the proposed method for the Mostly building and Building shadow are quite competitive. Therefore, the comparisons demonstrate the superior performance of the proposed NNCNet on the MUUFL dataset. The classification maps of the proposed NNCNet with / without pretraining are illustrated in Fig. \ref{fig_muufl_map}, it can be observed that the pretraining effectively improved the classification performance.

Table \ref{table_houston2018} illustrates the classification results of different methods on the Houston 2018 dataset. Compared to other methods, the proposed NNCNet achieves the best performance. Especially for `cars' and `paved parking lots', our method achieves 92.16\% and 95.25\%, which is far ahead of other methods. The classification maps are shown in Fig. \ref{fig_houston2018_map}. It can be seen that the results of other methods are not smooth enough for car classification, while the proposed NNCNet can depict the clear boundaries of cars and paved parking lots. It is evident that the proposed NNCNet has strong capabilities for fine-grained feature representation.

We find that the performance of the proposed NNCNet on the Houston 2013 dataset and Houston 2018 dataset far exceeds that on the Trento and MUUFL datasets. We believe it is due to the higher image resolution of both datasets (348$\times$1905 and 2384$\times$601 pixels). Therefore, the proposed NNCNet can exploit better feature representations on large dataset through contrastive learning. As a result, we believe that the proposed NNCNet could achieve better classification results in practical applications, in which more unlabeled data are available.

\subsection{Ablation Study}

To evaluate the effectiveness of different components in NNCNet, we conducted a series of ablation studies. The effectiveness of each proposed module for improving classification accuracy is verified through a series of ablation experiments, and the specific experimental results are listed in Table \ref{table_ablation}.

\textbf{Effectiveness of the Pretraining and Nearest Neighbor Learning}. We adopt a vanilla convolutional neural network without pretraining, bilinear attention fusion, and nearest neighbor contrastive learning as our baseline model. As illustrated in Table \ref{table_ablation}, compared with the baseline model, pretraining effectively improves classification performance to some extent on four datasets. It demonstrates that our pretraining scheme yields parameter initialization that can boost the classification accuracy.

We further examine our nearest neighbor-based contrastive learning scheme. As illustrated in Table \ref{table_ablation}, the model with nearest neighbor learning significantly boosts the classification performance. The reason is that the semantic similarities of neighborhood regions are taken into account, and the intermodal semantic alignments are enhanced.

To further demonstrate the effectiveness of the pretraining and nearest neighbor learning, we visualized the features before and after pretraining in Fig. \ref{fig_tsne}. We visualized the features without/with pretraining, together with the features in our final model, respectively. On the Houston 2013, Houston 2018 and Trento datasets, we found that after pretraining, the features of the same class distributed close to each other and the features of different classes moved far away from each other. It is evident that our unsupervised framework is effective on the Houston 2013 and Trento datasets. Furthermore, we observed that the features after pretraining do not improve significantly on the MUUFL dataset. The reason may be that there are more unlabeled data in the Houston 2013, Houston 2018 and Trento datasets. These unlabeled data play a critical role in contrastive learning. Therefore, the proposed contrastive learning framework performs better when more unlabeled data are available. It is more convenient in practical applications in which large amounts of unlabeled data are available.

\textbf{Number of Training Samples.} One of the advantages of self-supervised learning strategy is its excellent performance in handling small number of training samples. Therefore, we try to gradually reduce the number of samples during the training process, and the results are shown in Fig. \ref{fig_sam_num}. On the Houston 2013 dataset, when we use only 375 training samples (25 samples for each class), the OA value of the proposed method is 91.86 which is satisfying and encouraging. Furthermore, the model with pretraining consistently outperforms that without pretraining on four datasets when small training sets are used. It is evident that the contrastive learning strategy of the proposed NNCNet is especially effective for small training sets. Moreover, we observe that the performance gain of pretraining on the Houston 2013 and 2018 datasets is better than that on the Trento and MUUFL datasets. As mentioned before, there are more unlabeled data on the Houston 2013 and 2018 datasets. Therefore, the proposed nearest neighbor-based strategy can exploit rich feature representations on both datasets.

\textbf{Effectiveness of Data Augmentation}. The purpose of data augmentation is to enhance the differences between positive and negative samples as a way to facilitate the training of the encoder. In the proposed NNCNet, we use four data augmentation schemes, including RandomResizedCrop, RandomHorizontalFlip, RandomVerticalFlip and RandomGaussianNoise. The corresponding results are shown in Fig. \ref{fig_dataaug}. We found that RandomResizedCrop is the key to data augmentation. Since the image patch is cropped into 11$\times$11 pixels, if the scale is set too small, the semantic information would easily be damaged. Therefore, in our implementations, the scale is set to (0.7, 1).

\subsection{Parameter Sensitivity}

\textbf{Minimum Spatial Distance between Positive and Negative Samples}. In order to prevent too much similarity between positive and negative samples, we define a minimum distance $s$ between them (i.e. the distances between positive and negative samples need to be greater than $s$). The results are shown in Fig. \ref{fig_queue_size}(a). In our implementations, the size of each sample is $11\times11$ pixels. The classification performance improved slightly when $4\leqslant s \leqslant 12$. It is beneficial to use a large distance to increase the difference between positive and negative samples. Therefore, in our implementations, $s$ is set to 12. 

\textbf{Size of the Negative Key Dictionary.} Fig. \ref{fig_queue_size}(b) shows the effect of negative key dictionary size on the classification performance. The experiments show that a larger dictionary size will have a positive effect on pretraining, and it is consistent with our previous assumptions. We believe that the proposed method works better when more unlabeled data are available.

\textbf{Key Encoder Update Speed}. We tested different key encoder update speeds $r$ during pretraining. The experimental results are shown in Fig. \ref{fig_queue_size}(c). We find that the best classification performance is achieved when $r$ is set to 0.9.

\textbf{Effectiveness of 3D Convolution}. Inspired by HybridSN \cite{roy2019hybridsn}, we first use PCA for  channel dimensionality reduction. Then, 3D and 2D convolutions are combined for feature extraction. To verify the effectiveness of 3D convolution, we design a network in which the 3D convolutions are replaced with 2D convolutions (``w/o Conv2d" in Fig. \ref{fig_3dconv}). The experimental results are shown in Fig. \ref{fig_3dconv}. We found that 3D convolution can improve the classification performance to some extent. Although PCA disturbs the spectral continuity of the hyperspectral data, we argue that 3D convolution can still generate more discriminative feature maps from the spectral dimensions than 2D convolution. These discriminative features generated by 3D convolution can boost the classification performance.

\section{Conclusions and Future Work}

In this paper, we propose a self-supervised NNCNet model to tackle the HSI and LiDAR joint classification problem. Specifically, we integrate a nearest neighbor-based data augmentation scheme into the contrastive learning framework. Semantic similarities among neighborhood regions are exploited. The intermodal semantic alignments can be captured more accurately. In addition, we proposed a bilinear attention fusion module that can capture second-order feature interactions between HSI and LiDAR data. Therefore, the module improves the contextual representation of multisource data effectively. Extensive experiments on Houston 2013, Trento, MUUFL and Houston 2018 datasets have demonstrated the superiority of our model to a wide range of state-of-the-art methods.

In the future, we aim to explicitly explore the semantic and spatial relations between HSI and LiDAR data. In addition, we will explore how to further enhance the feature interactions between HSI and LiDAR data.

\begin{IEEEbiography}[{\includegraphics[width=1in,height=1.25in,clip,keepaspectratio]{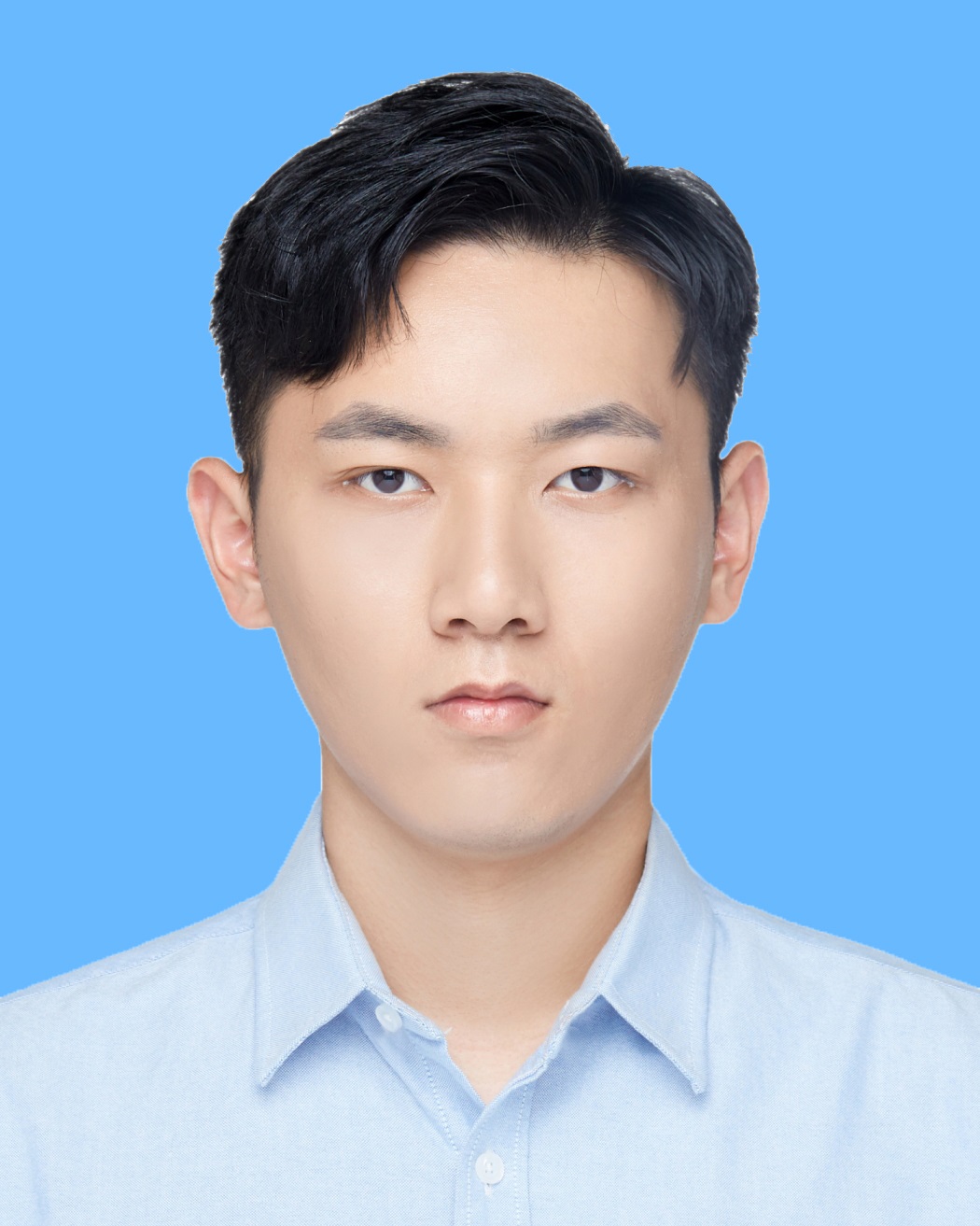}}]{Meng Wang}
received the B.Sc. degree in computer science from Jinan University, Jinan, China, in 2020. He is currently pursuing the M.Sc. degree in computer science and applied remote sensing with the School of Information Science and Technology, Ocean University of China, Qingdao, China.

His current research interests include computer vision and remote sensing image processing.

\end{IEEEbiography}

\begin{IEEEbiography}[{\includegraphics[width=1in,height=1.25in,clip,keepaspectratio]{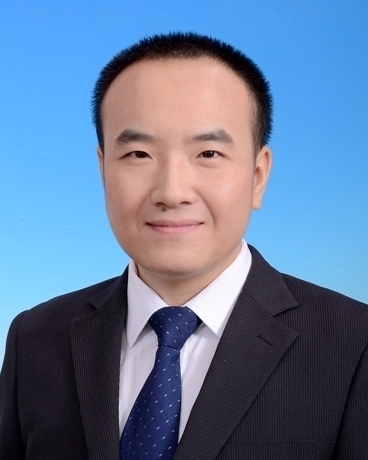}}]{Feng Gao} (Member, IEEE)
received the B.Sc degree in software engineering from Chongqing University, Chongqing, China, in 2008, and the Ph.D. degree in computer science and technology from Beihang University, Beijing, China, in 2015.

He is currently an Associate Professor with the School of Information Science and Engineering, Ocean University of China. His research interests include remote sensing image analysis, pattern recognition and machine learning.

\end{IEEEbiography}

\begin{IEEEbiography}[{\includegraphics[width=1in,height=1.25in,clip,keepaspectratio]{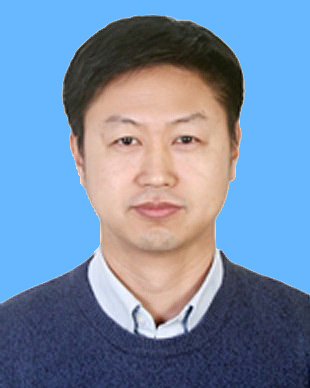}}]{Junyu Dong}
 (Member, IEEE) received the B.Sc. and M.Sc. degrees from the Department of Applied Mathematics, Ocean University of China, Qingdao, China, in 1993 and 1999, respectively, and the Ph.D. degree in image processing from the Department of Computer Science, Heriot-Watt University, Edinburgh, United Kingdom, in 2003.

He is currently a Professor and Dean with the School of Computer Science and Technology, Ocean University of China. His research interests include visual information analysis and understanding, machine learning and underwater image processing.
\end{IEEEbiography}

\begin{IEEEbiography}[{\includegraphics[width=1in,height=1.25in,clip,keepaspectratio]{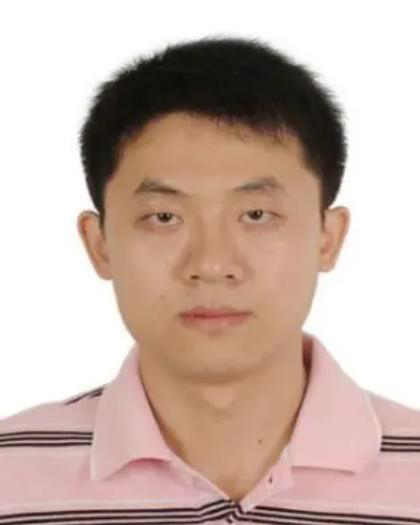}}]{Heng-Chao Li} (Senior Member, IEEE) received the B.Sc. and M.Sc. degrees from Southwest Jiaotong University, Chengdu, China, in 2001 and 2004,
respectively, and the Ph.D. degree from the Graduate University of Chinese Academy of Sciences, Beijing, China, in 2008. 

He is currently a Full Professor with the School of Information Science and Technology, Southwest Jiaotong University. His research interests include statistical analysis of synthetic aperture radar (SAR) images, remote sensing image processing, and pattern recognition.

Dr. Li is an Editorial Board Member of the \textit{Journal of Southwest Jiaotong University} and \textit{Journal of Radars}. He is an Associate Editor of the \textsc{IEEE Journal of Selected Topics in Applied Earth Observation and Remote Sensing}.

\end{IEEEbiography}

\begin{IEEEbiography}[{\includegraphics[width=1in,height=1.25in,clip,keepaspectratio]{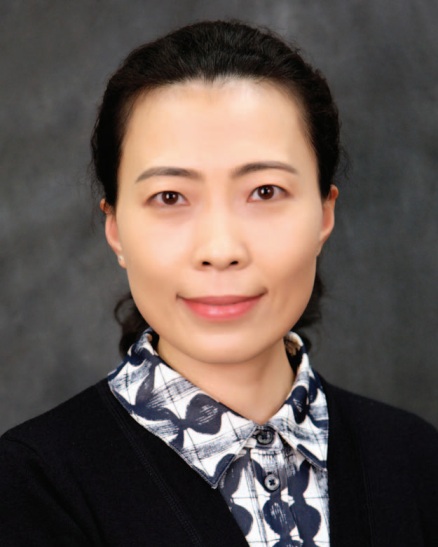}}]{Qian Du}
(Fellow, IEEE) received the Ph.D. degree in electrical engineering from the University of Maryland at Baltimore, Baltimore, MD, USA, in 2000.

She is currently the Bobby Shackouls Professor with the Department of Electrical and Computer Engineering, Mississippi State University, Starkville, MS, USA. Her research interests include hyperspectral remote sensing image analysis and applications, and machine learning.

Dr. Du was the recipient of the 2010 Best Reviewer Award from the IEEE Geoscience and Remote Sensing Society (GRSS). She was a Co-Chair for the Data Fusion Technical Committee of the IEEE GRSS from 2009 to 2013, the Chair for the Remote Sensing and Mapping Technical Committee of International Association for Pattern Recognition from 2010 to 2014, and the General Chair for the Fourth IEEE GRSS Workshop on Hyperspectral Image and Signal Processing: Evolution in Remote Sensing held at Shanghai, China, in 2012. She was an Associate Editor
for the \textsc{Pattern Recognition}, and \textsc{IEEE Transactions on Geoscience and Remote Sensing}. From
2016 to 2020, she was the Editor-in-Chief of the \textsc{IEEE Journal of Selected Topics in Applied Earth Observation and Remote Sensing}. She is currently a member of the IEEE Periodicals Review and Advisory Committee and SPIE Publications Committee. She is a Fellow of SPIE-International Society for Optics and Photonics (SPIE).

\end{IEEEbiography}

\bibliography{source}
\bibliographystyle{IEEEtran}

\end{document}